\documentclass[aps,prb,twocolumn,showpacs,superscriptaddress]{revtex4}
\usepackage{graphicx}
\usepackage{color}

\newcommand{\EqLabel}[1]{\label{#1}}

\begin{document}

\newcounter{mycounter} \newenvironment{noindlist}
{\begin{list}{\arabic{mycounter}.~~}{\usecounter{mycounter}
\labelsep=0em \labelwidth=0em \leftmargin=0em \itemindent=0em}}
{\end{list}}

\title{The polaron paradigm: a dual coupling effective band model}

\author{Dominic J. J. Marchand} \affiliation{Department of Physics and
Astronomy, University of British Columbia, Vancouver, BC, Canada,
V6T~1Z1}

\author{Philip C. E. Stamp} \affiliation{Department of Physics and
Astronomy, University of British Columbia, Vancouver, BC, Canada,
V6T~1Z1}
\affiliation{Pacific Institute of Theoretical Physics, University
of British Columbia, Vancouver, BC, Canada, V6T~1Z1}

\author{Mona Berciu} \affiliation{Department of Physics and Astronomy,
University of British Columbia, Vancouver, BC, Canada, V6T~1Z1}

\begin{abstract}

Non-diagonal couplings to a bosonic bath completely change polaronic
dynamics, from the usual diagonally-coupled paradigm of
smoothly-varying properties. We study, using analytic and numerical
methods, a model having both diagonal Holstein and non-diagonal
Su-Schrieffer-Heeger (SSH) couplings. The critical coupling
found previously in the pure SSH model, at which the $k=0$ effective
mass diverges, now becomes a transition line in the coupling constant
plane - the form of the line depends on the adiabaticity
parameter. Detailed results are given for the quasiparticle and ground
state properties, over a wide range of couplings and adiabaticity
ratios. The new paradigm involves a destabilization, at the transition
line, of the simple Holstein polaron to one with a finite ground-state
momentum, but with everywhere a continuously evolving band shape. No
'self-trapping transition' exists in any of these models. The physics
may be understood entirely in terms of competition between different
hopping terms in a simple renormalized effective band theory. The
possibility of further transitions is suggested by the results.

\end{abstract}

 \maketitle

\section{Introduction}

An electron moving in a solid (ordered or otherwise) polarizes its
surroundings, and dresses itself with bosonic excitations which
include electronic spin and charge fluctuations, electronic orbital
fluctuations, and phonons. The result is a {\it polaron}; the
best-known example is an electron dressed by a combination of optical
and acoustic phonons. The phonon-dressed polaron is of enduring
interest for two main reasons:

\vspace{1mm}

(i) Polarons are believed to play a key role in determining the physics of many
 materials; and

\vspace{1mm}

(ii) It provides a target for any theory of quasiparticles which tries
to span the weak, intermediate, and strong coupling regimes. This has
given the subject an interesting history (briefly described in section
II), closely connected to many of the important developments in
many-body physics and quantum field theory.

\vspace{1mm}

For most of this history, attention has focused on a specific kind of
"diagonal"  coupling between the density of
  electrons (this is a diagonal operator in a real space formulation)
  and phonon displacements; the latter can be local or longer-range,
  as reviewed in section II. Such models are now well understood, and
have led to a "polaron paradigm" in which diagonal interactions, no
matter what their form, do not lead to any kind of sharp transition in
the polaron properties as the coupling strength is increased (eg., the
 effective mass increases smoothly with coupling, up to
infinite coupling strength).

However one may also have non-diagonal couplings to the bath, only
acting when the particle hops between sites (which in the continuum
limit couple to the particle momentum). We now know that such
couplings lead to results very different from the classical polaron
paradigm. This was first shown in Ref. \onlinecite{PRL} for a
specific model with a non-diagonal coupling (the SSH model); a sharp
transition was found between the behavior at weak coupling and that
at strong coupling. As we shall see here, this transition persists
even when we add diagonal interactions to the non-diagonal ones.

These results appear to be rather general, indicating that the old
polaron paradigm needs to be replaced by a rather different one
incorporating the interplay between diagonal and non-diagonal
couplings. This is what we do in this paper. The results are not just
of theoretical interest - they will force a substantial re-evaluation
of our picture of polarons in many physical systems.

The plan of the paper is as follows. Section II discusses models
containing both diagonal and non-diagonal terms, and briefly reviews
previous work. Section III describes our method,  the
  'Bold Diagrammatic Monte Carlo' (BDMC) and the momentum average (MA)
  approximation. The former method is also discussed in Appendix A,
which is an integral part of the paper. Sections IV and V give the
main results; Section IV discusses the results when only non-diagonal
SSH terms are present, highlighting the transition in the polaronic
properties. Section V adds diagonal terms, and shows how the
transition is influenced by these terms. Section VI summarizes the new
picture of the polaron that comes out of these results, in the form of
an effective band theory leading to a new kind of polaron paradigm -
readers looking for a quick summary should go to this section, which
also discusses experimental applications.

\section{The model}

We would like to study a model which brings out the main features of
both diagonal and non-diagonal couplings without being too
complicated. In this section we first develop the model, and then
discuss a few simple key features it possesses, at the same time
recalling some of the main results found over the years.

\subsection{Derivation of the dual coupling model}

We begin by formulating the dual coupling model in a fairly general
way. In a site basis we start from a Hamiltonian
\begin{eqnarray}
\hat{\mathcal{H}}_o \;=\; - \sum_{ij} t_{ij}( \{ b_{\lambda} \})
c_{i}^{\dagger} c_{j} &+& \sum_i \epsilon_i ( \{ b_{\lambda} \})
c_{i}^{\dagger} c_{i} \nonumber \\ &+& \sum_{\lambda} \omega_{\lambda}
b_{\lambda}^{\dagger} b_{\lambda}
 \label{Ho-site}
\end{eqnarray}
describing the hopping, with amplitude $t_{ij}$, of a particle between
sites $i$ and $j$ of a lattice, in the presence of a bosonic bath
having a single branch of excitations of frequency $\omega_{\lambda}$
(here, $\lambda$ labels the quantum numbers of the bosons; for
example, we can choose $\lambda = \{ {\bf q}, \mu \}$, where ${\bf q}$
is a momentum and $\mu$ a polarization; or we can choose $\lambda = \{
i,\mu \}$, with $i$ a site index again). Both the on-site energy
$\epsilon_i$ and the hopping matrix element $t_{ij}$ are modulated by
(i.e., are functionals of) the boson variables $b_{\lambda},
b_{\lambda}^{\dagger}$. If we are dealing with lattice phonons, the
modulation is through the site displacement operators
$\hat{x}_{\lambda} = \sqrt{\hbar\over 2M\omega_{\lambda}}
\left(b_{\lambda} +b_{\lambda}^\dagger\right)$, where $M$ is the ionic
mass. This generates the diagonal couplings (modulations of
$\epsilon_i$) and non-diagonal couplings (modulations of $t_{ij}$).

The form of (\ref{Ho-site}) is quite general - the lattice may take
any form (ordered or disordered), as may the various coupling
terms. To say more we must specify how $\epsilon_i$ and $t_{ij}$
depend on the bosonic variables. This dependence can take many forms
for different physical systems. In what follows, we will assume that
we are dealing with a periodic crystalline lattice, and that the
bosons in questions are lattice phonons; crystal momentum is then a
good quantum number.

Let us expand the Hamiltonian (\ref{Ho-site}) in terms of the phonon
variables $\{ b_{\lambda} \}$. Then the diagonal interaction terms are
produced by expanding the on-site energy:
\begin{eqnarray}
\epsilon_i = \epsilon_o &+& \sum_{\lambda} U^{(1)}_i(\lambda)
(b_{\lambda} +b_{\lambda}^\dagger) \nonumber \\ &+& \sum_{\lambda
  \lambda'} U^{(2)}_i(\lambda, \lambda') (b_{\lambda}
+b_{\lambda}^\dagger)(b_{\lambda'} +b_{\lambda'}^\dagger) + ...
 \label{diag}
\end{eqnarray}
where $U^{(1)}_i(\lambda)$ is the 1-phonon diagonal coupling,
$U^{(2)}_i(\lambda, \lambda')$ the 2-phonon diagonal coupling, and so
on. This diagonal term arises simply from the polarization of the
background lattice by the particle - for electrons, it is
essentially a Coulomb interaction effect.

The non-diagonal couplings are produced by expanding $t_{ij}(
\{b_{\lambda} \})$ in phonon variables. This is more subtle, since the
hopping matrix element arises from electron tunneling, and so depends
exponentially on the phonon displacement. Often this dependence is
written as
\begin{equation}
t_{ij}( \{ b_{\lambda} \}) = t_o \exp \left[ - \sum_{\lambda} { V_{ij}
    (\lambda) \over \omega_{\lambda}} (b_{\lambda}
  +b_{\lambda}^\dagger) \right]
 \label{t-ph}
\end{equation}
where $V_{ij} (\lambda)$ has the dimension of energy; but this form
generates only one part of the multi-phonon interaction terms (which
can also appear directly in the exponent).  However since we are only
interested in the linear term, we expand (\ref{t-ph}) to
linear order in phonon variables, so that
\begin{equation}
t_{ij}( \{ b_{\lambda} \}) = t_o \left[1 - \sum_{\lambda} { V_{ij}
    (\lambda) \over \omega_{\lambda}} (b_{\lambda}
  +b_{\lambda}^\dagger) \right]
 \label{t-Ph-Lin}
\end{equation}
The physical meaning of this 'Peierls' term is obvious from the
derivation - the phonons modulate the distance between the lattice
ions, and this modulates the tunneling amplitude of the electron
between sites. As we shall see, the effective range of the $t_{ij}$, once it is
renormalized by the coupling to phonons, can be significantly greater
than a single lattice spacing.

In this paper we will deal exclusively with ordered lattices. We then
Fourier transform the starting Hamiltonian (\ref{Ho-site}); and as
further restrictions, we will (i) keep only the linear coupling to
phonons; (ii) assume a single optical phonon branch, with polarization
index suppressed and frequency $\omega_{\bf q}$; and (iii) assume the
lattice has no impurities, defects, surfaces, or other inhomogeneities
that break translational invariance. The Hamiltonian describing such
problems then has the general structure:
\begin{eqnarray}
  \label{Ham}
\hat{\mathcal{H}} &=&\sum_{k}\epsilon_k c_k^\dagger c_k + \sum_q
\omega_q b_q^\dagger b_q \nonumber \\ & + & \frac{1}{\sqrt{N}}
\sum_{k,q} V(k,q) \; c_{k-q}^\dagger c_k \big( b_{q}^\dagger +b_{-q}
\big),
 \label{H-pol}
\end{eqnarray}
where we suppress the electronic spin variable. Here $V(k,q)$ is the
sum of the Fourier-transformed diagonal and non-diagonal terms,
$c_k^\dagger, b_q^\dagger$ are electron and phonon creation operators,
$k$ is the electron momentum and momentum sums are over the first
Brillouin zone. We assume $N$ lattice sites and let $N \rightarrow
\infty$. The elimination of both acoustic phonons and higher-order
phonon couplings restricts the applicability of (\ref{H-pol}) in the
real world. In section VI we return to  these
  approximations and their consequences.

Splitting $V(k,q)$ into its diagonal and non-diagonal parts, we have
\begin{equation}
g({ k}, { q}) = g_1({ q}) + g_2({ k}, { q})
 \label{g12}
\end{equation}
where the diagonal coupling $g_1({ q})$ depends only on the phonon
momentum ${ q}$, while the non-diagonal coupling $g_2({ k,q})$
depends explicitly on the particle momentum ${ k}$.

\vspace{2mm}

{\it The Dual "H/SSH" model}: So far these results are purely formal.
In most of the rest of this paper we wish to obtain explicit results
for the polaron properties in a model which combines two specific
forms for the diagonal and non-diagonal couplings. We therefore
consider a 1-dimensional chain with lattice constant $a_o = 1$, and a
bare hopping term $t_{ij} \rightarrow t_o \delta_{i\pm 1,j}$, i.e.,
constant nearest-neighbor hopping only. We choose the phonons to be a
set of Einstein phonons of frequency $\Omega_o$. The non-interacting
part of ${\cal H}$ in (\ref{H-pol}) is then
\begin{equation}
{\cal H}_o \;=\; - 2t_o \sum_k \cos k \; c_k^\dagger c_k + \sum_q
\Omega_o b_q^\dagger b_q
 \label{H-o}
\end{equation}
giving two characteristic energy scales, $t_o$ and $\Omega_o$.

For the diagonal electron-phonon coupling we use a simple on-site
Holstein coupling,\cite{Holstein} so that
\begin{equation}
g_1(q) \rightarrow g_o
 \label{V1-g}
\end{equation}
with no dependence on momentum at all.  For the non-diagonal coupling
we use the so-called Su-Schrieffer-Heeger (SSH)
form,\cite{PhysRevLett.42.1698,RevModPhys.60.781,BLB1,BLB2} which 
describes the modulation of the hopping term, viz.,
\begin{equation}
t_{i,i+1} \rightarrow \; t_o - \alpha_o (x_{i+1} - x_{i})
 \label{hopC}
\end{equation}
which in momentum space gives a coupling
\begin{equation}
g_2(k,q) = \frac{2i}{\sqrt{N}} \alpha_o \left[ \sin(k-q) - \sin
  k\right]
 \label{V2-a}
\end{equation}
and we have introduced 2 new energy scales, $g_o$ and $\alpha_o$.

We see that if we scale everything in terms of $t_o$, there are 3
independent parameters in this "H/SSH" model. The adiabaticity
parameter is:
\begin{equation}
\Lambda_o = \Omega_o/(4t_o),
 \label{Lambda}
\end{equation}
so that when $\Lambda_o \ll 1$, the phonon dynamics is considered as
slow (and the electron dynamics is fast). One can also define
dimensionless parameters $g_o/t_o$ and $2\alpha_o/t_o$, with
dimensionless ratio $2\alpha_o/g_o$. However, a better understanding
of the physics is obtained by defining
\begin{eqnarray}
\lambda_H &=& {g_o^2\over 2t_o \Omega_o} \;\;\;\;\;\;(diagonal)
\\ \lambda_{SSH} &=& {2\alpha_o^2 \over t_o \Omega_o}
\;\;\;\;\;\;(non-diagonal)
 \label{lambda}
\end{eqnarray}
whose ratio is now $(2\alpha_o/g_o)^2$. Along with $\Lambda_o$, these
dimensionless parameters determine the behavior of this model at zero
temperature.

This is the simplest model one can look at for a polaronic
system with both diagonal and non-diagonal couplings. It is something
of a toy model and it ignores acoustic phonons. Nevertheless it
is good enough to reveal the main features of an entirely new
behavior for the polaron, one quite different from the polaronic
paradigm described above. How much these general features will persist
in more general models is a question we address at the end of the
paper.

\subsection{Basic features of the model}

 A great deal is known about the effect of the coupling $g_1$ on
 polaron dynamics, much less about $g_2$, and still less about the
 effect of a combination of the two. The main results that have so far
 been established are as follows.

\vspace{2mm}

(i) {\it Diagonal terms}: As noted above, many models with different
diagonal couplings have been studied. The most famous examples are the
Holstein model\cite{Holstein} (where, as just noted, $g_1({ q})
\rightarrow g_o$, a constant) and the Fr\"ohlich model,
\cite{Frohlich} where the form of $g_1({ q})$ depends on the spatial
dimension and on whether we deal with a lattice or a continuous medium
model). Other examples include the Rashba-Pekar model,\cite{RaPe}
breathing-mode (BM) models,\cite{bmo} and various models with displacement
potential coupling.\cite{Mahan} A variety of theoretical methods to
study them were developed in the period from 1940 to 1970, including
perturbation expansions in the interaction, \cite{lang:firsov:63}
semiclassical approximations \cite{landau33,Toyozawa}, and path
integral techniques.\cite{Feynman} The main aim was to understand the
behavior of the polaron as the strength of the diagonal coupling
$g_1({ q})$ was increased. Indeed, the problem was (and still is)
regarded as providing a key test of non-perturbative methods, and thus
of interest well beyond solid-state physics. More recently, other
methods have been developed, notably the Dynamical Mean-Field Theory
(DMFT)\cite{DMFT} and the Momentum Average (MA)
approximations.\cite{MA,MAh,MAq} The predictions of all of these
methods can now be checked by a wide variety of numerical techniques,
including exact diagonalization, variational methods, various types of
Quantum Monte Carlo simulations, and in one dimension, by Density
Matrix Renormalization Group (DMRG) methods. The literature on all
this is now enormous.\cite{reviews}

These diagonal models all have certain broad features in common. For
weak coupling, there is typically very little phonon dressing, and the
polaron properties are only slightly renormalized. This weak coupling
regime is sometimes called the {\em large polaron} regime (although
that term may be more appropriate for continuous models). As the coupling is
increased, there is a crossover to the {\em small polaron} regime,
where a robust polarization cloud is formed, and the electronic
properties are strongly renormalized. The effective mass then
increases rapidly with coupling (eg., exponentially fast in the
Holstein model). For a long time there was significant confusion over
the idea that in this strong-coupling regime the polaron might become
self-trapped (i.e. localized), although it is now clear that such
self-trapping is impossible in a clean system for any finite coupling.
Disorder in real systems can of course localize polarons, particularly
when they are heavy; but this is a different effect. Another related,
and long-standing controversy, was over whether there is a sharp
transition or a smooth crossover between the two regimes. This
question was settled by the work of Gerlach and L\"owen
\cite{RevModPhys.63.63} who showed that for diagonal models having
gapped bosonic modes, sharp transitions in the polaronic properties
are impossible; all physical quantities must vary smoothly with
coupling strength. We note that the so-called 'self-trapping
transition', which has in the past often been asserted to exist at
some finite value of the coupling, is actually typically an artefact
of numerical approximations, arising when the effective bandwidth is
decreasing rapidly with increasing coupling constant.

Thus we see that from all of this work a general consensus has emerged for the
diagonal coupling model, both on the essential physics and on the detailed
quantitative picture. This is the "{\it polaron paradigm}";
as conventionally understood, it involves a smooth crossover between
weak and strong
coupling, and between large and small polaron behavior, with no sharp
transition of any kind; and for a clean system, no "self-trapping" or
localization of the polaron on any particular site.

\vspace{2mm}

(ii) {\it Non-Diagonal Terms}: Much less work has been done on the
non-diagonally coupled model, and most of this has been in the context
of applications to systems like polyacetylene (the SSH model
\cite{PhysRevLett.42.1698, RevModPhys.60.781, RiceMele}) and other
polyacenes, \cite{polyacene,rubrene} as well as excimers,
\cite{excimer} MX chains,\cite{MX} and the cuprate
superconductors.\cite{nagaosa04} Most theoretical studies of the
single polaron in such models have been fairly recent, and we can separate them into
two categories.

First, several studies have argued that in dual coupling models,
having both diagonal and non-diagonal interactions, one can discern a
'self-trapping' transition line in the 2-dimensional plane of the 2
couplings. Evidence cited for this has typically come from variational
analyses, using either a Toyozawa ansatz \cite{sun09} or a
'global-local' ansatz \cite{PhysRevB.79.165105}. There are also
perturbative analyses \cite{PhysRevB.56.4484,PhysRevB.66.012303,Mars},
supplemented by exact diagonalization studies \cite{PhysRevB.56.4484}
on very small (4-site) lattices, that have claimed evidence for a
crossover (not a transition) between large and small polaron regimes
as a function of a non-diagonal SSH coupling (note  that
Refs. \onlinecite{PhysRevB.66.012303} and \onlinecite{Mars} have non-diagonal coupling to
acoustic rather than optical phonons).

Second, a few papers have looked at the entanglement between the
polaronic particle and the phonon bath as a function of the coupling
constants - this was first done for the Holstein model,\cite{zhao04}
and then for the SSH model.\cite{stojan08} In the pure Holstein model
the entanglement increases smoothly during the crossover between the
large and small polaron limits (this is described as a cliff-like
transition by Zhao et al., \cite{zhao04} but in fact their results show
only a smooth change in the entanglement). Stojanovi\ifmmode
\acute{c}\else \'{c}\fi{} and Vanevi\ifmmode \acute{c}\else \'{c}\fi{}
then found,  in the SSH model, a non-analyticity in
the entanglement as a function of the SSH coupling,\cite{stojan08} but they then
argued that this did not signal any kind of transition in the
polaronic properties (in particular, no change in the ground state),
but rather a loss of coherence of the polaron.

In sections IV and V we will demonstrate that the physics is very
different from that proposed by these earlier analyses. In fact there
really is a sharp transition line in the dual coupling parameter
plane, but it has nothing to do with any kind of 'phase transition' or
even a crossover between large and small polarons, or to a
self-trapped state; nor is it in the same part of the parameter space
as the transitions claimed in the previous work. Instead, it is
associated with a change in the ground state, coming from a continuous
evolution of the band structure as a function of the non-diagonal
coupling. The $k=0$ polaron effective mass actually diverges along the
transition line - but this is simply because of the shift of the ground
state momentum; the bandwidth is finite even at the transition. Thus,
the polaron is mobile everywhere in the parameter plane, even on the
critical line. The crossover between large and small polarons turns
out to be entirely associated with the diagonal part of the coupling -
the non-diagonal coupling plays no role in this.  Sections IV-V gives
full details of the results, and section VI describes the new picture
that emerges from them.

\section{Methods: Perturbative and Numerical}

One reason we are confident in the accuracy of the results discussed
herein is that we have been able to benchmark them against results
found with the extremely powerful diagrammatic Monte Carlo (DMC)
method invented by Prokof'ev, Svistunov and
Tupitsyn.\cite{prokofev-1998-87} In the current work we have augmented
this method with an improved version, the Bold Diagrammatic Monte
Carlo (BDMC) method, and we have also used the much faster
Momentum Averaging (MA) approximation. In what follows we
describe how they are applied to the present problem - technical
details for the BDMC method are relegated to the Appendix. We also say
a little about perturbation theory - this turns out to work well only
in the extreme anti-adiabatic regime when $\Lambda_o \gg 1$.

In Ref. \onlinecite{PRL}, results were reported for the SSH model
using the above methods and also the Limited Phonon Basis Exact
Diagonalization (LPBED) method.\cite{PhysRevB.80.195104} The agreement
between the LPBED, DMC, BDMC and MA results was found to be excellent,
and we expect this to also be true for the present dual-coupling model
because all these methods also work well for the Holstein model.

\subsection{Perturbation theory}

For perturbative work, we write $\Sigma(k,\omega) = \sum_l \Sigma_l
(k,\omega)$, where $\Sigma_l (k,\omega)$ is the sum of all self-energy
graphs containing $l$ internal boson lines. Both the
Rayleigh-Schrodinger and Wigner-Brillouin versions of perturbation
theory are then applied.  In Rayleigh-Schr\"odinger (RS) perturbation
theory it is well-known \cite{rspt_frolich} that for polaron models
with Einstein phonons and $\Omega_o < 4t_o$, the first order
contribution to $E_k$ exhibits an unphysical maximum at some momentum
$k_m >0$, and an unphysical divergent energy at larger
momenta. Because of this, we can only calculate the ground state
energy and the effective mass if the ground-state is located at
$k=0$. At second order, the unphysical maximum is eliminated and so is
the divergent negative ground state energy, but for $\Omega_o < 4t_o$
one finds instead a positive divergent energy at larger momentum.

In Wigner-Brillouin (WB) perturbation theory, where the polaron energy
is approximated using the implicit equation $E_k = \epsilon_k + \mbox{Re}
\Sigma(k,E_k)$, no singular values for $E_k$ are found anywhere in
the Brillouin zone. WB is also useful as a consistency check for any
numerical code: if electron lines are not dressed, and only
contributions of up to two phonon lines are allowed, the resulting
self-energy must converge to $\Sigma_2(k,\omega)$. Since the polaron
energy $E_k$ is the lowest pole of the Green's function $G(k,\omega) =
[\omega - \epsilon_k -\Sigma(k,\omega)]^{-1}$, with these restrictions
any numerical results must coincide with WB.

\subsection{Momentum Average (MA) approximation }

The MA technique is a non-perturbative approximation which has been
applied successfully to a number of polaron problems.
\cite{MA,MAh,MAq,PhysRevB.82.085116,MAca} It sums all diagrams in the
self-energy expansion, up to exponentially small contributions which
are discarded. The MA self-energy is then expressed in closed form as
a continued fraction, and evaluates this very efficiently. MA can be
systematically improved\cite{MAh} so that its convergence can be
assessed.  It can also be cast in variational terms, as keeping
contributions only from processes consistent with a certain structure
of the phonon cloud. In its simplest, original formulation (already
surprisingly accurate for the Holstein model), the phonon cloud can be
arbitrarily far from the electron and contain arbitrarily many
phonons, but its size is limited to one site.\cite{MAh,MAvar} For the
Edwards model\cite{Edwards} -- another model describing
boson-modulated hopping -- it was shown that to correctly describe the
polaron dynamics, the minimum extent of the cloud is over three
adjacent sites (again, an arbitrary number of bosons is allowed at
each site, and the particle can be arbitrarily far away from the
cloud).\cite{PhysRevB.82.085116}

The MA results presented here have been generated using a
straightforward generalization of the work in
Ref. \onlinecite{PhysRevB.82.085116}. On general grounds, we expect it
to be accurate if $\Lambda_o$ is not too small; if phonons are
energetically expensive, having them spread over many sites becomes
very unlikely. Of course, one can generalize MA to allow for more
extended clouds, but the calculations become rather cumbersome. As
shown below, the MA results produced here are quantitatively quite
accurate as long as $\Lambda_o \ge 0.1$ or so. Note that we are using
here what is technically called a MA$^{(0)}$ method, which does not
allow additional phonons to exist away from the polaron cloud. As a
result, it cannot describe the polaron $+$ one-phonon continuum,
expected to appear at an energy $\Omega_o$ above the polaron
ground-state energy. MA properly captures this continuum for
MA$^{(1)}$ or higher level approximations, which include these states
amongst those kept in the variational space.\cite{MAh} However, here
we are concerned with the polaron band which lies below this
continuum, and this MA version is already sufficient to describe it
accurately, and to allow us to efficiently evaluate various trends.

\subsection{Bold Diagrammatic Monte Carlo (BDMC)}

The BDMC method \cite{PhysRevLett.99.250201, PhysRevB.77.020408,
  PhysRevB.77.125101} is an elaboration of the earlier Diagrammatic
Monte Carlo (DMC) algorithm.\cite{prokofev-1998-87} Both belong to a
category of unbiased computational methods, in which no assumptions
are made about the solution or the type of diagrams that contribute
most to the final answer. Instead, all the diagrams are summed using
stochastic sampling, and the result is exact within the statistical
error bars. The DMC method was further developed to study the
Fr\"ohlich polaron, \cite{PhysRevLett.81.2514, PhysRevB.62.6317} the
Holstein polaron\cite{Macridin2003} and the spin polaron,
\cite{PhysRevB.64.033101} amongst others. Recently, it was also
extended to many-polaron problems.\cite{Mnew}

The method is formulated in imaginary time, where all diagrams are
real functions of the internal lines.  For diagonal coupling models,
with $V(k,q) = g_1(q)$, this removes any sign problem because in any
self-energy diagram all emitted bosons are reabsorbed, so the vertices
contribute a positive factor $g_1(q) g_1(-q) = |g_1(q)|^2>0$
irrespective of the topology of the diagram. In this case their sum is
convergent, and success is guaranteed.

For models with non-diagonal couplings $g_2(k,q)$, however, this is no
longer necessarily true. Hermiticity of the Hamiltonian guarantees
that $g_2(k',-q)=g_2^*(k'+q,q)$, so if we consider the vertex product
$g_2(k,q) g_2(k',-q)$ for a phonon line with momentum $q$, its
contribution is positive only if $k'=k-q$, {\em i.e.}, if this phonon
line is not crossed by other phonon lines. At first sight, it looks as
if these contributions are complex, not real; however, this is not the
case because of symmetries. For example, for our dual-coupling H/SSH
model, a purely imaginary contribution comes from diagrams where an
odd number of phonon lines are either emitted by a Holstein process
and absorbed by a SSH process, or vice-versa. One can check that the
contribution of such a diagram is canceled by that of its
time-reversed pair. Only diagrams with an even number of both Holstein
and of SSH vertexes survive. These have real expressions, but diagrams
with crossed phonon lines can now have either sign.

Thus we see that for $\alpha_o \ne 0$, our model has a sign problem. It is not a
severe problem because at least all the non-crossed diagrams are
positive, but it makes it desirable to perform
partial summations of diagrams, which may further mitigate it. This
idea gave rise to the BDMC algorithm, in
which free propagator lines are replaced by ``bold'' lines which
already include some of the self-energy contributions. This is done in
an iterative fashion, with proper care to avoid double counting of
diagrams.

To the best of our knowledge, our results reported in
Ref. \onlinecite{PRL} and here are the first BDMC results for a single
polaron problem; in the Appendix we explain in more detail how these
calculations were done.

\section{Results: Holstein and SSH models}

Before looking at the results when we have both couplings in the
system, we examine each model separately. The physics of the diagonal
Holstein model is well-known - our purpose in discussing it here is to
demonstrate the accuracy of the methods. We then examine the SSH model using
the same methods, and discuss the physics around the critical coupling
where the transition takes place, and the physical mechanism which
causes the transition.

\subsection{Pure Holstein model}

\begin{figure}[t]
\includegraphics[width=\columnwidth]{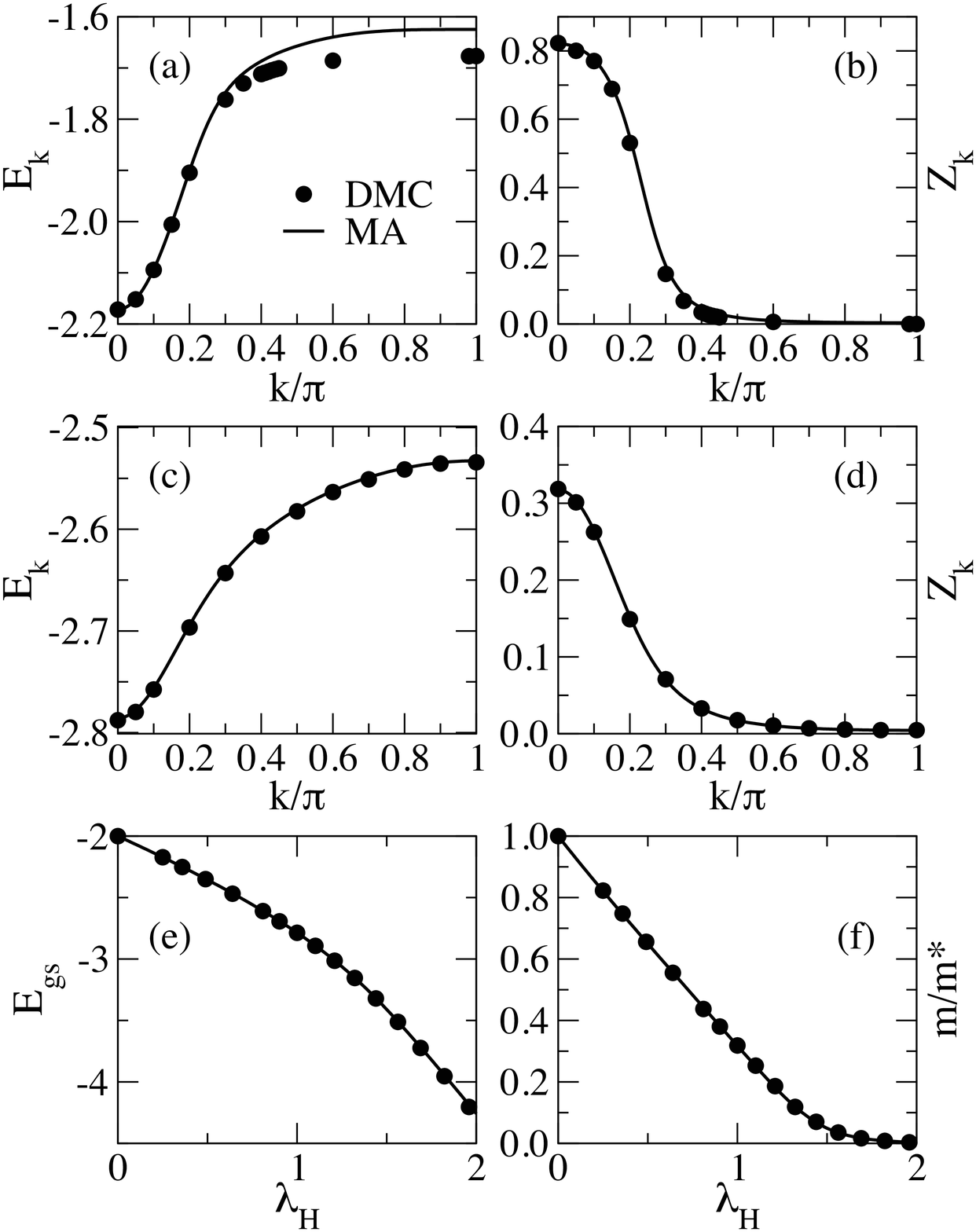}
\caption{\label{fig1} Results for the 1D Holstein model, using DMC
  (symbols) and MA (lines). (a) Polaron energy $E_k$ and (b)  its
  $qp$ weight $Z_k$, for $\lambda_H=0.25$; (c) polaron energy
  $E_k$ and (d) its $qp$ weight $Z_k$, for $\lambda_H=1.0$; (e)
  polaron ground-state energy $E_{gs}$ and (f) its inverse effective
  mass $m/m^*(k=0)$ vs. $\lambda_H$ at $k=0$. In all cases, $t_o=1$,
  $\Lambda_o \equiv \Omega_o/(4t_o)=0.125$, and $\lambda_{SSH}=0$. DMC
  data are from Ref. \onlinecite{Macridin2003}.}
\end{figure}

The standard properties of interest in the diagonally-coupled models
like the Holstein model are (i) the polaronic properties, notably the
quasiparticle dispersion relationship $E_k$ and the related
quasiparticle weight $Z_k$, along with the polaronic effective mass
$m^*$, as functions of both quasiparticle momentum and electron-phonon
coupling strength; and (ii) the energy $E_{gs}$ of the polaronic
ground state, i.e., the minimum quasiparticle energy, as a function of
the coupling strength.

In what follows we describe one set of results for this model, to see
how well the different numerical methods work in this well-understood
case; this also serves to illustrate the main features of the standard
polaron paradigm. Both MA and DMC data are given (there is no
difference between DMC and BDMC results; BDMC is simply
computationally more efficient).

Fig. \ref{fig1} shows different quasiparticle quantities for the
Holstein polaron. To be specific, we assume an adiabacity parameter
$\Lambda_o =0.125$. Panels (a) and (b) show the polaron dispersion
$E_k$ and quasiparticle ($qp$) weight $Z_k$ in half of the Brillouin
zone, in the weak coupling regime (for $\lambda_H=0.25$). Here the MA
results overestimate the bandwidth -- as already discussed above, this
is because this is a MA$^{(0)}$-type method, which does not properly
account for the location of the polaron+one-phonon continuum (higher
level MA approximations fix this problem\cite{MAh}). As the Holstein
coupling $\lambda_H$ increases, the polaron bandwidth decreases below
$\Omega_o$ and the agreement becomes excellent everywhere, as shown in
panels (c) and (d) for $\lambda_H=1$. The agreement improves even
further for larger $\lambda_H$ (not shown), since any flavor of MA
becomes exact \cite{MA} in the limit $\lambda_H \rightarrow \infty$.
The decreasing bandwidth corresponds here to an increasing effective
mass $m^*(k)$, where we define this mass in the usual way as
\begin{equation}
m^*(k) = \bigg[\frac{\partial^2 E_k}{\partial k^2}\bigg]^{-1}
\end{equation}
The inverse of this effective mass is shown in (f) for $k=0$. We
notice how rapidly the bandwidth and corresponding inverse effective
mass decrease for $\lambda_H > 1$ (thus, for $\lambda_H = 1$, we see
from (c) that the bandwidth $\sim 0.25t_o$, whereas for $\lambda_H =
2$, it is $\sim 0.005 t_o$). However, we also note that there is no
sudden collapse or discontinuity of any kind in the effective mass as
a function of $\lambda_H$.

In panel (e) we look at the ground state energy, and confirm that the
MA and DMC methods agree with each other; ground-state properties are
described accurately by MA for all couplings $\lambda_H$.  No
qualitative changes are expected for these properties, if $\Lambda_o$
is increased. Note that the accuracy of the MA results is improved,
especially at intermediary couplings, over that shown in Fig. 10 of
Ref. \onlinecite{MA}. This is because here we allow for a polaron
cloud extending over up to three adjacent sites, whereas there we only
considered a one-site cloud.

These results illustrate well the standard polaron paradigm. As
coupling increases, the bandwidth decreases monotonically with a
corresponding increase in the effective mass. The basic shape of the
dispersion relation is unchanged, with the usual monotonic increase of
energy with momentum, qualitatively similar to that of the bare
particle. All properties vary smoothly with $\lambda_H$; as expected,
there is no trace of any kind of transition or non-analyticity in
either the quasiparticle properties or the
energy,\cite{RevModPhys.63.63} as functions of either the coupling
strength or the momentum. Although details will change if we change
the momentum dependence of the coupling $g(q)$, the polaron behavior is qualitatively similar.

\subsection{Pure SSH model}

\begin{figure}[t]
 \includegraphics[width=\columnwidth]{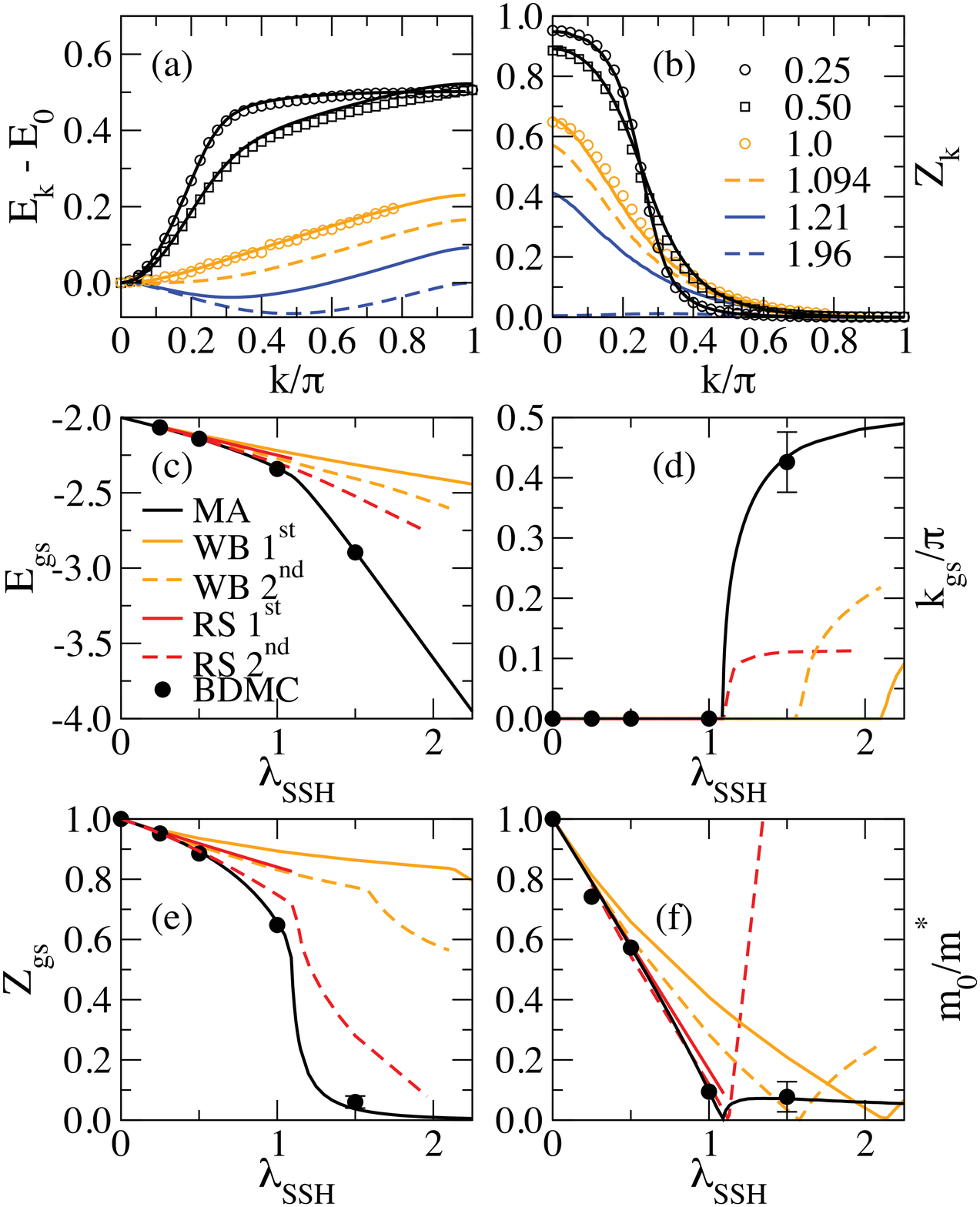}
\caption{\label{fig2} (color online) Results for the 1D SSH model with
  $\Lambda_o \equiv \Omega_o/(4t_o) =0.125$ from BDMC, MA, and 1$^{st}$ and
  2$^{nd}$ order RS and WB perturbation theory. (a) Shifted polaron
  energy $E_k-E_0$ and (b) its $qp$ weight $Z_k$, for $\lambda_{SSH}=
  0.25, 0.5, 1, 1.094, 1.21, 1.96$, with BDMC (symbols) and MA
  (lines). (c) Polaron ground-state energy $E_{gs}$, (d) the
  ground-state momentum $k_{gs}$, (e) the ground-state $qp$ weight
  $Z_{gs}$ and (f) the inverse effective mass $m/m^*(k=k_{gs})$, all
  as functions of $\lambda_{SSH}$.  Panels (c)-(f) share the same
  colour code described in the legend of panel (c). For BDMC results,
  error bars are shown only when they are larger than the size of the
  symbols. In all cases, $t_o=1$ and $\lambda_H=0$. }
\end{figure}

In the SSH model we need to look at the quasiparticle properties a
little differently. This is because the key feature is the sharp
transition which we find at a critical coupling $\lambda_{SSH}^*
(\Lambda_o)$; this transition changes the way we must characterize
the polaron quasiparticle. In addition to
the usual quantities $E_k, Z_k$, and $m^*$, we also need to look at
the ground state momentum $k_{gs}$, which changes rather quickly once
$\lambda_{SSH}> \lambda_{SSH}^*$; and the ground state energy,
which of course now depends on $k_{gs}$.

(i) {\it Numerical results}: To see what happens, we begin in
Fig. \ref{fig2} with SSH results for the same value $\Lambda_o=0.125$ as
above, so we can directly compare with the Holstein result of
Fig. \ref{fig1}. Panels (a) and (b) show the polaron energy $E_k$
(shifted so that all curves start at zero) and $qp$ weight $Z_k$. For
weak couplings, the curves look rather similar to those of the
Holstein model -- the dispersion is  monotonic and it flattens
out just below the continuum (whose energy is again slightly
overestimated by MA). However, for medium and strong couplings the
results are very different. The dispersion does not simply become
flatter; instead it changes its shape  so that the ground-state is no longer at
$k_{gs}=0$. In (c) and (d) we see clearly that the transition involves
a non-analytic dependence of both the ground-state energy
$E_{gs}(k_{gs})$ and the ground-state momentum $k_{gs}$ on the
coupling constant, with discontinuities in all derivatives of these
quantities with respect to $\lambda_{SSH}$ at $\lambda_{SSH} =
\lambda_{SSH}^*$. Notice from (c) and (d) that, even though
perturbation theory gives a very poor description of $E_{gs}$ and
$k_{gs}$ as functions of $\lambda_{SSH}$, nevertheless a 2$^{nd}$-order
Rayleigh-Schrodinger calculation does capture the correct position of
the singularity.

This pattern is repeated in calculations of $Z_k$ and the effective
mass $m^*(k)$ as functions of $\lambda_{SSH}$, provided we evaluate
these at the bottom of the band ($k=k_{gs}$). Indeed we see from (f)
that the effective mass $m^*(k_{gs}) \rightarrow \infty$ as
$\lambda_{SSH} \rightarrow \lambda_{SSH}^*$; this is accompanied by a
divergence in the derivative $\partial Z(k_{gs})/\partial
\lambda_{SSH}$ (see (e)). However, the divergence in $m^*(k_{gs})$ is
simply due to the inflection point in $E_k$ at $k=0$ when
$\lambda_{SSH} = \lambda^*_{SSH}$, and therefore does not reflect a
collapse of the bandwidth, as obvious from (a) (the dispersion curve
at the transition point $\lambda^*_{SSH} = 1.094$ has a finite
bandwidth). In fact, even for large couplings the bandwidth is still
considerable. Thus, for $\lambda_{SSH}=1.96$, the bandwidth is $\sim
0.1t_o$ as opposed to $\sim 0.007t_o$ for the Holstein model at
similar coupling $\lambda_H=1.96$. We also notice that well above the
transition, the effective mass varies rather slowly; in fact
$m^*(k_{gs}) \approx 10-20 m$ even for large $\lambda_{SSH}$.

\begin{figure}[t]
\includegraphics[width=\columnwidth]{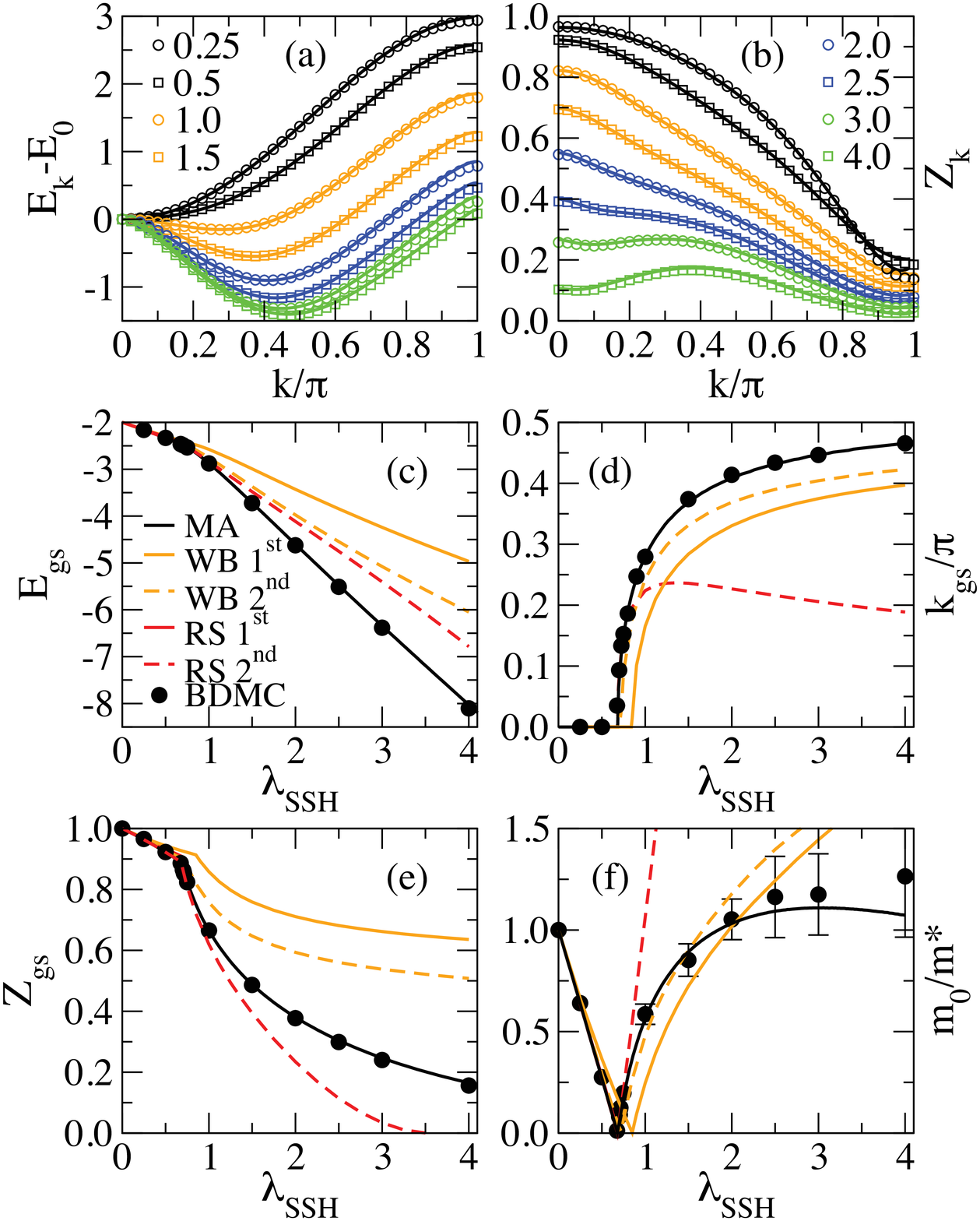}
\caption{\label{fig3} (color online) Results for the 1D SSH model with
  $\Lambda_o = 0.75$. We show the same quantities as in Fig. \ref{fig2}; in (a) and (b) we
show results for $\lambda_{SSH}$
ranging from 0.25 to 4.0.  }
\end{figure}

We see that we are dealing here with a sharp transition between a
weak- and a strong-coupling regime. The weak-coupling
regime has a single non-degenerate ground state located at
$k_{gs}=0$. Since the system is inversion symmetric, the
strong-coupling regime is doubly degenerate, with ground state
wave-vectors $\pm k_{gs}\ne 0$.  As emphasized in
Ref. \onlinecite{PRL}, the transition arises from a non-analyticity in
the function $E_{gs}(\lambda_{SSH})$ at $\lambda_{SSH} =
\lambda^*_{SSH}$. The non-analyticity is clearly not associated with
any localization or self-trapping of the polaron; nor is it a quantum
phase transition. This is because although we work at $T=0$, we are
only treating the single-electron case. A phase transition
involves the cooperative behavior of a macroscopic number of degrees
of freedom, which is simply impossible in the one-electron limit,
where only a finite and rather small average number of phonons are
associated with the single polaron cloud. It is of course 
possible that the transition we have found may evolve into a critical
phase transition at a finite particle concentration; however we cannot
support such claims based on the current data.

\begin{figure}[t]
 \includegraphics[width=\columnwidth]{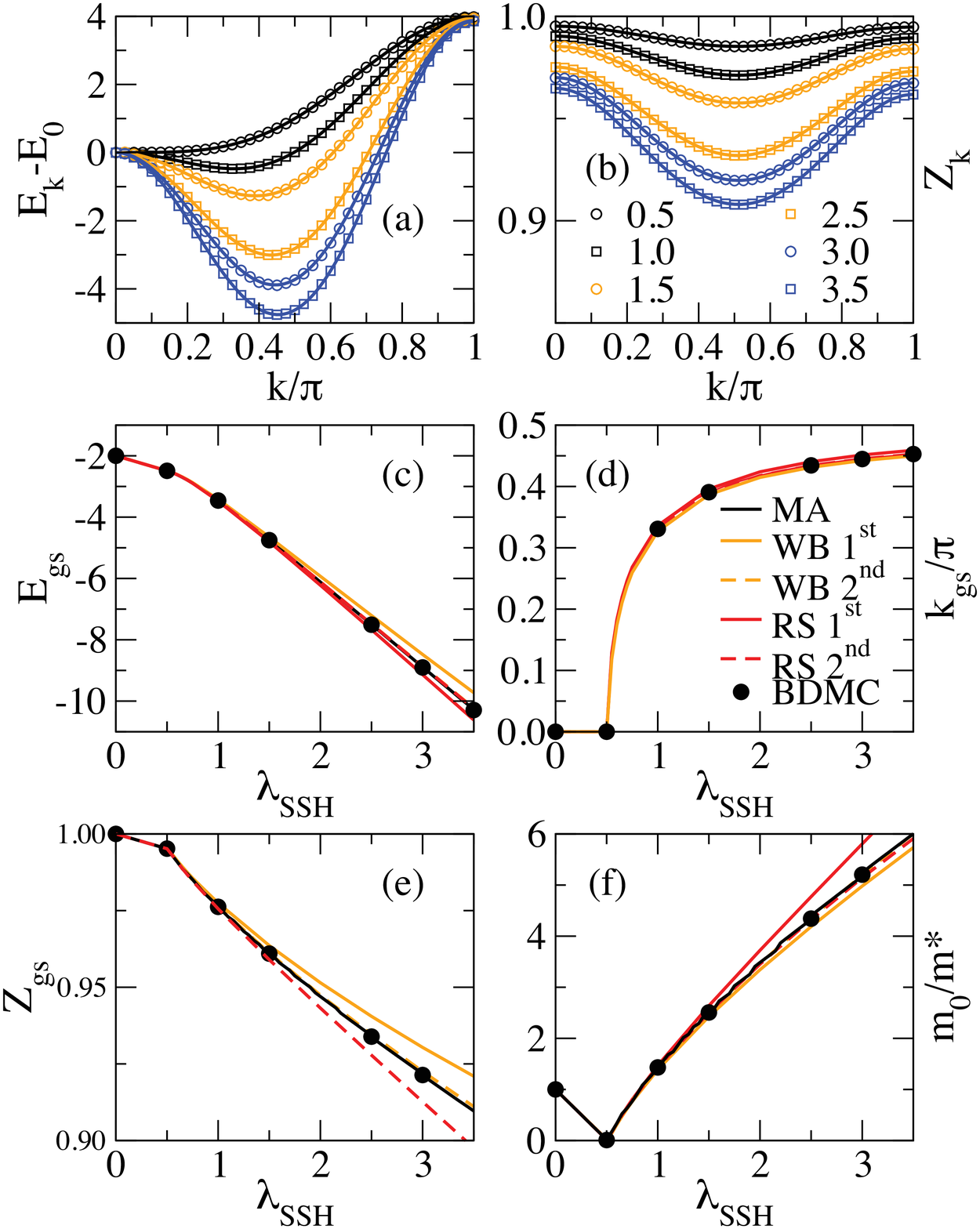}
\caption{\label{fig4} (color online) Results for the 1D SSH model,  with
  $\Lambda_o  = 25$.  We show the same quantities as in Fig. \ref{fig2}; in (a) and (b) we
show results for $\lambda_{SSH}$
ranging from 0.5 to 3.5. }
\end{figure}

It remains to check how the results vary with the adiabaticity
parameter $\Lambda_o$. In Figs. \ref{fig3} and \ref{fig4} we show all
of the same results as those in Fig. \ref{fig2}, but now for
$\Lambda_o = 0.75$ and $25$. The main observations we can make here are:

(a) The basic behavior of the dispersion relation as a function of
$\lambda_{SSH}$ is just as before, but as $\Lambda_o$ increases, the
bandwidth  steadily increases, and the effective mass
$m^*(k_{gs})$ decreases - indeed, quite remarkably, for large
$\Lambda_o$ the polaron is actually lighter than the bare band particle
except in the region near the critical coupling, where its mass
diverges. The transition is signaled as before by non-analytic
behavior in all functions at $\lambda_{SSH} = \lambda^*_{SSH}$; the
basic behavior of $k_{gs}$ and $E(k_{gs})$ as functions of
$\lambda_{SSH}$ is similar to that for small $\Lambda_o$. Notice how
well perturbation theory works in the extreme anti-adiabatic limit
$\Lambda_o \gg 1$; the reason for this is discussed below.

\begin{figure}[t]
\includegraphics[width=\columnwidth]{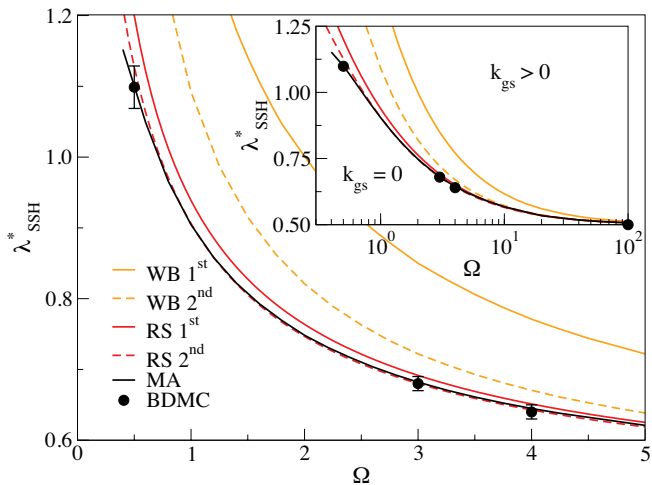}
\caption{\label{fig5} (color online) Transition line $\lambda^*_{SSH}$
  vs. $\Omega_o$ in the pure SSH model, separating the
  weak-coupling regime, with its non-degenerate ground state at
  $k_{gs}=0$, from the strong-coupling regime with its doubly
  degenerate ground-state at $\pm k_{gs}\ne 0$. The inset shows the
  same data over a wider range of phonon energies. Results from BDMC,
  MA, and perturbational RS and WB of both $1^{st}$ and $2^{nd}$ order
  are shown. Other parameters are $\lambda_H=0, t_o=1$. }
\end{figure}

(b) As $\Lambda_o$ increases, the $qp$ weight is less and less
affected by the non-diagonal coupling to the phonons, and is not much
smaller than the free particle result $Z_k = 1$; for $\Lambda_o \gg 1$,
we see that $Z_k$ remains considerable even at very large
couplings. This is of course the exact opposite of what happens in any
diagonally coupled model, no matter what the form of $g(q)$.

(c) We also see that as $\Lambda_o$ increases, the transition coupling
$\lambda^*_{SSH}$ slowly decreases. In Fig. \ref{fig5}, we show this
behavior in more detail, by drawing the "critical line"
$\lambda^*_{SSH}(\Lambda_o)$ as a function of the adiabaticity ratio,
which separates the weak-coupling and strong-coupling regimes, and
along which the effective mass diverges. All methods, even 1st-order
perturbation theory, work well in the anti-adiabatic limit $\Lambda_o
\gg 1$, where we see that $\lambda^*_{SSH} \rightarrow 0.5$, a result
which we explain below.

On the other hand we do not show any numerical results for the
adiabatic limit $\Lambda_o \rightarrow 0$ in Fig. \ref{fig5}, because
all these methods become inaccurate here - note the large error bars
for $\Lambda_o \ll 1$. This is because the average number of phonons
in the polaron cloud increases rapidly with decreasing $\Lambda_o$,
and the convergence of the BDMC code is then drastically slowed. MA
results do converge, however the three-site cloud assumption becomes
questionable in this limit. As noted in Ref. \onlinecite{PRL}, LPBED
results based on a five-site cloud restriction suggest that as
$\Omega_o \rightarrow 0$, $\lambda^*_{SSH}$ actually reaches a maximum
and then decreases. Thus what actually does happen for small
$\Lambda_o$ is an open question. This region can be studied
using other numerical methods\cite{adia} that have already been
successfully applied to the Holstein model in the adiabatic limit.

\vspace{2mm}

(ii) {\it Physics of the pure SSH polaron}: The behavior discussed
above is most easily understood if we begin from the anti-adiabatic
limit, $\Lambda_o \rightarrow \infty$, where perturbation theory can
be applied (for a preliminary discussion of this limit see
Ref. \onlinecite{PRL}). Consider the first order contribution to the
polaron self-energy, viz.:
$$ \Sigma_1(k,\omega) = {1\over N} \sum_{q}^{}
\frac{|V(k,q)|^2}{\omega+i\eta - \epsilon_{k-q} -\Omega_o}.
$$ Since the range of $\epsilon_k$ is just the bandwidth, i.e., $\sim
\mathcal{O}(t_o)$, it follows that when $\Omega_o \gg t_o, \omega$, the
denominator is $\sim -\Omega_o$. The integral can then be carried
out, and we find:
$$ \lim_{\Lambda_o \rightarrow \infty} \Sigma_1(k,\omega) = -t_o
\lambda_{SSH}[2- \cos(2k)]
$$ Higher-order diagrams contributing to $\Sigma_l(k,\omega)$, i.e.,
containing $l$ internal phonon lines, scale as $(1/\Omega_o)^{2l-1}$,
and can thus be ignored for sufficiently large $\Lambda_o$ (note that
we deal here with an asymptotic series - the number of
diagrams at level $l$ is $ \sim (2l-1)!!$, which for $l > \Lambda_o$
outweighs the factor $(1/\Omega_o)^{2l-1}$; but for $\Lambda_o \gg 1$,
this simply means that low-order perturbation theory gives very
accurate results, \cite{divergentS} unless we go to very long times).
It then follows that the polaron energy $E_k = \epsilon_k +
\Sigma(k,E_k)$ in this limit is well approximated by:
\begin{equation}
\EqLabel{asymp} E_k = -2t_o\lambda_{SSH} -2t_o \cos(k) + t_o
\lambda_{SSH} \cos(2k).
\end{equation}
In other words, apart from an overall shift, the energy contains both
nearest-neighbor and second-nearest-neighbor hopping terms. The
former favors a $k=0$ ground state, while the latter, because of its
unusual sign, favors a ground state at $k=\pm {\pi\over 2}$, together
with a ``folded'' dispersion. For small $\lambda_{SSH}$ the former
term dominates, but if $\lambda_{SSH}$ is sufficiently large, the
behavior is controlled by the phonon-mediated second-nearest-neighbor 
term. The transition takes place at the inflection
point where the effective mass diverges, i.e., when:
$$ \frac{m}{m^*(k=0)} = -2t_o[1-2\lambda_{SSH}] \; \rightarrow 0
$$ explaining why, in Fig. \ref{fig5}, $\lim_{\Omega_o \rightarrow
  \infty} \lambda^*_{SSH}=0.5.$ These considerations are confirmed by
the data shown in Fig. \ref{fig4} for $\Omega_o=100t_o$. As expected,
here all perturbative results agree well with the BDMC and MA results.

In physical terms, what has happened here is that the phonon-modulated
hopping has opened a new ``channel'' for particle motion. Direct
hopping of a particle away from its phonon cloud has an exponentially
small probability when phonons are energetically
costly. This is why in diagonal coupling models, with ${V}(k,q) =
g(q)$, the effective hopping amplitude (and hence the inverse
effective mass) is exponentially suppressed. However, in a
non-diagonal hopping model with ${V}(k,q) = V(k,q)$, the particle can
move phonons along as it hops. In the SSH model discussed here,
the particle can hop to the neighboring site while creating a phonon
on that site, and then hop one site further while absorbing this same
phonon. This gives rise to an effective second nearest-neighbor
hopping, i.e., we now have an effective low-energy band Hamiltonian for
the polaron of form
\begin{equation}
E_k = t_1 \cos k \;+\; t_2 \cos 2k
 \label{Ek2t}
\end{equation}
where $t_1$ is the renormalized value of the bare $t_o$,
and $t_2$ the effective 2$^{nd}$ nearest-neighbor hopping term. The phonons have not
gone away, of course, but for this lowest polaron band $\mbox{Im}
\Sigma(k,\omega)=0$, since there are no lower states for the polaron
to decay to. For the SSH model the transition at
$\lambda_{SSH} = \lambda^*_{SSH}$ occurs because $t_2$ has a negative
sign - the physical reason for this is that the phonon increases one
bond length, but decreases the other (cf. discussion in
Ref. \onlinecite{PRL}).

More complicated scenarios are found in other $g(k,q)$ models; however
the underlying idea is the same. For example, in the Edwards
model,\cite{Edwards,PhysRevB.82.085116,Mono} the particle first creates a string of three
consecutive bosons and then goes back and removes them, again
resulting in effective second nearest neighbor hopping (this time
with a positive sign, so that there is no transition). Similar results
are obtained in  $t-J_z$
models.\cite{Holger2} Here, the particle goes twice around closed
Trugman loops.\cite{Trloops} On the first circuit it creates a string
of bosons, which are all removed on the second circuit. This gives
rise to both effective second- and third-nearest neighbor hopping.

Now consider what happens when $\Omega_o$ decreases. The one-phonon
mechanism for the transition is then supplemented by processes
involving more and more phonons, which will create not only
second-nearest neighbor hoppings, but also longer-range hopping terms
(and lead to increasing disagreement between low-order perturbation
theory and the BDMC and MA results). Note that if the phonons modulate
only the nearest neighbor hopping -- as in the SSH model -- then only
terms $t_{2n}$ can be generated dynamically, because there must be an
even number of hops to absorb all emitted phonons, and so the carrier
cannot end up an odd number of sites away from its initial
location without changing the number of bosons. The $t_{2n+1}$ terms are either renormalized bare terms,
like $t_1$, or are generated by mixing with the continuum as
$\Omega_o$ decreases and the band is flattened. Of course, in models
where phonons modulate longer-range hopping and/or coupling is beyond
linear, any $t_n$ can be generated dynamically.

\begin{figure}[t]
\includegraphics[width=\columnwidth]{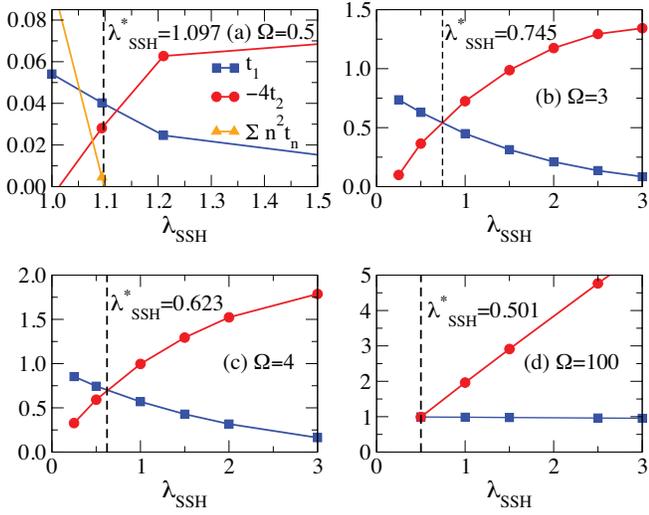}
\caption{\label{fig6} (color online) Effective hopping constants $t_1$
  (squares)
  and $-4 t_2$  (circles)
  in the SSH model, obtained from a fit of $E_k$ as described by
  Eq. (\ref{fit}), for (a) $\Omega_o = 0.5t_o$, (b) $\Omega_o = 3t_o$,
  (c) $\Omega_o = 4t_o$ and (d) $\Omega_o =100t_o$. Triangles, shown
  only in (a), represents $\sum_{n=1}^{5}
  n^2 t_n$. See text for more details.}
\end{figure}

To investigate this dynamical mechanism further we fit the polaron
dispersion $E_k$, obtained from BDMC, to a functional form
\begin{equation}
\EqLabel{fit} E_k = - \sum_{n=0}^{5} 2 t_n \cos(nk),
 \label{t-n}
\end{equation}
up to fifth nearest-neighbor hopping. The results are
shown in Fig. \ref{fig6}. For $\Omega_o=100t_o$, in panel (d), the
fits agree with Eq. (\ref{asymp}): $t_1 \approx t_o$, $t_2
\approx -t_o \lambda_{SSH}/2$, and the higher $t_{n\ge3}$ are
too small to show. As $\Lambda_o$ decreases, a fit to $E_k$ with only
first and second nearest neighbor hopping remains acceptable,
although  the values of $t_1$ and $t_2$ deviate
significantly from the asymptotic limits (compare panels (b) and
(c)). The transition occurs when $t_1=-4t_2$; the crossing of these
lines agrees well with $\lambda^*_{SSH}$ (dashed line), for these
larger values of $\Lambda_o$.

However if we now go to $\Omega_o =0.5$, shown in panel (a), then
longer-range hopping must be included  to obtain reasonable
fits. For simplicity, we continue to plot only $t_1$ and $-4t_2$.  The
$t_1$ and $-4t_2$ lines cross at a value above $\lambda^*_{SSH}$,
because of the $n\ge 3$ hopping terms. The role of these terms can
be seen directly if we compute the inverse effective mass $1/m^*(k=0)
= \; \sum_{n=1}^{5} n^2 t_n$. This value is shown by triangles, and it
indeed vanishes at $\lambda^*_{SSH}$.

Clearly as $\Lambda_o$ decreases further it becomes impractical to try
and unravel all the higher-order terms. This is why we cannot explain
all the details of the transition in this limit, nor can we predict
what happens as $\Lambda_o \rightarrow 0$ based on these methods and results.

Let us now summarize what we have found for the polaron with purely
off-diagonal SSH coupling. Just as we saw that the physics of the
diagonally-coupled polaron could be summarized in one key feature,
viz., the smooth crossover from weak to strong coupling, with no
intervening transition, we see that the non-diagonally coupled
polaron's behavior also turns on one key feature, viz., the range and
sign of the phonon-mediated long-range hopping terms, which may lead
to a sharp transition in polaron properties. The physics here is
intrinsically more complicated because one can imagine a wide variety
of scenarios, depending on the relative sizes and signs of the
different terms. This will be particularly true in the quasi-adiabatic
regime when $\Lambda_o$ is small, where we do not rule out a whole
sequence of transitions as the bare off-diagonal coupling
increases. In the simple SSH model studied here, there is only one
transition in the regime we have studied, where we assume that
$\Lambda_o$ is not too small - the reason for this was discussed
above. But clearly these results constitute just the tip of the
iceberg as far as possible different behaviors are concerned, and it
will be interesting to study different off-diagonally coupled models,
having different forms for $g(k,q)$.

\section{Dual-coupling model - combined Holstein and SSH couplings}

In any real polaronic system there will generically be both diagonal and
non-diagonal couplings between the electrons and the phonons (or
whatever other kind of bosonic excitation is involved). Thus it is
crucial to know how the competing (and very different) effects coming
from each coupling end up working together. In what follows we
investigate a combined "Holstein + SSH" model, with both $\lambda_H$
and $\lambda_{SSH}$ finite. Note that even though Eqs. (\ref{V1-g})
and (\ref{V2-a}) suggest that the results depend on the sign of
$g_o/\alpha_o$, this sign is in fact irrelevant. This is because, as
we noted when discussing the sign problem, all contributions with odd
powers in either $g_o$ and/or $\alpha_o$ cancel out. Thus the
parameters $\lambda_H$ and $\lambda_{SSH}$ are sufficient to fully
characterize the couplings in this model.

Before discussing our results, it is worthwhile trying to guess what
might be the result of combining the couplings. If we take the
'effective band' idea seriously, then one can argue that (i) the main
effect of the diagonal coupling is to simply reduce the bandwidth of
the "$\cos k$" bare band, i.e., to severely renormalize the
nearest-neighbor hopping $t_o$ down to $t_1$; and (ii) the main
effect of the non-diagonal coupling is to create longer-range hopping
terms of the  form $t_{2n} \cos 2nk$. Thus everything will now depend on
the relative strength of these terms - if we imagine first putting in
the non-diagonal term, to create the long-range hoppings, and then add
the diagonal term, we expect that not only will $t_o$ be renormalized
down to $t_1$, but that there will also be a reduction of the higher
$t_n$, for $n\ge 2$. Note that if the higher $t_n$ are
renormalized down by the same factor as the renormalization of $t_o$
to $t_1$, the critical line $\lambda^*_{SSH}$ will be unchanged as a
function of $\Lambda_o$: there will only be an overall weakening of
all hopping amplitudes. However one can also imagine a scenario where
the higher $t_n$ are less strongly reduced than $t_o$ - the critical
coupling $\lambda^*_{SSH}$ will then be reduced from its value when
there is no diagonal coupling. Conversely, if the $t_n$ are more
strongly reduced, we expect $\lambda^*_{SSH}$ to increase. Which
scenario will be exhibited is not {\it a priori} clear, nor is it
clear how these results will depend on the adiabaticity
ratio.

\begin{figure}[t]
\includegraphics[width=\columnwidth]{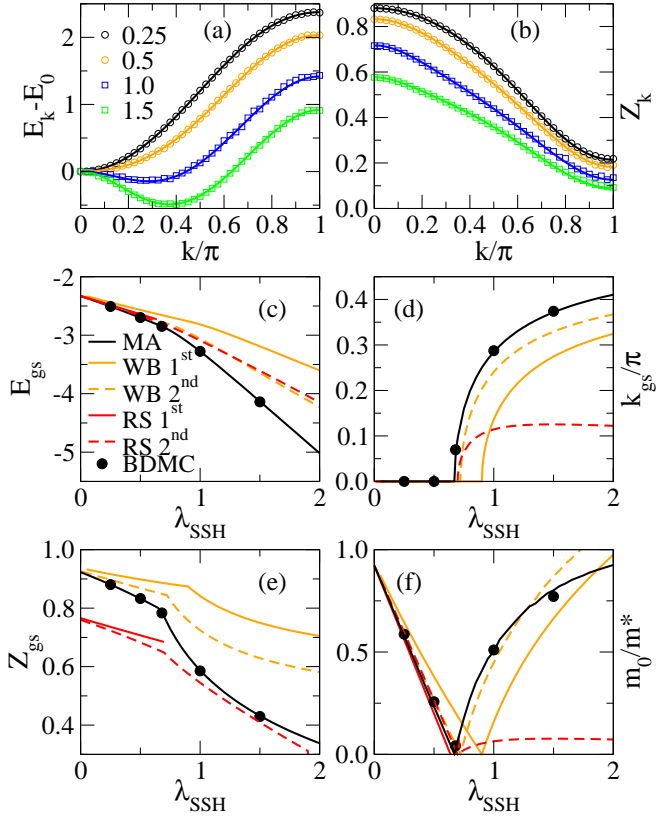}
\caption{\label{fig7} (color online) Polaron properties for the combined
  H/SSH model. We show the same plots  as those shown in
  Fig. \ref{fig2} for the pure SSH
  model, but now with a finite diagonal coupling $\lambda_H=0.25$. We
  choose $\Lambda_o=0.75, t_o=1$. In (a) and (b), the values of
  $\lambda_{SSH}$ are shown in the legend.}
\end{figure}

\begin{figure}[t]
\includegraphics[width=\columnwidth]{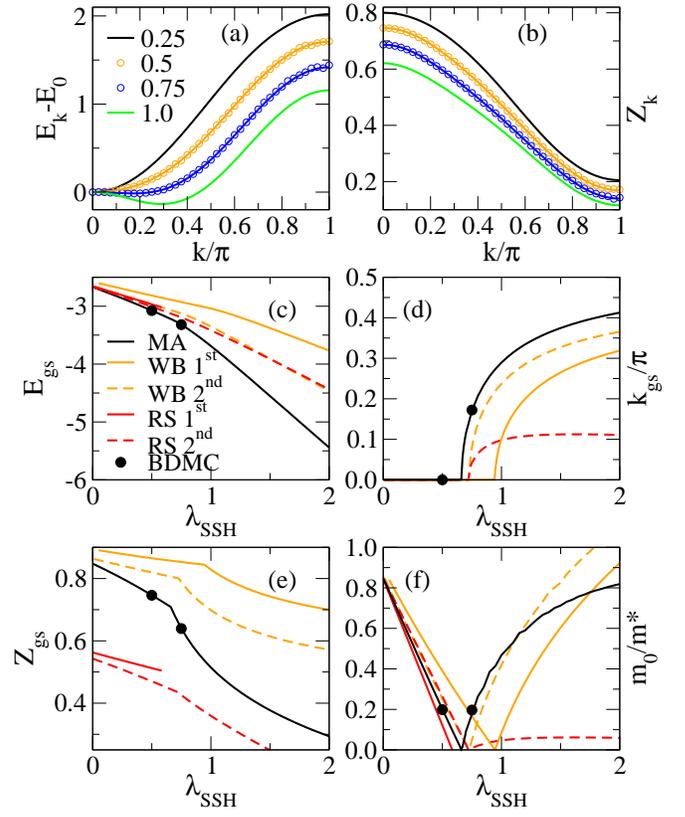}
\caption{\label{fig8} (color online) The same plots as in
  Fig. \ref{fig7}, but now for $\lambda_H=0.5$. As before,
  $\Lambda_o=0.75, t_o=1$. In (a) and (b), the values of $\lambda_{SSH}$
  are shown in the legend. }
\end{figure}

What is clear is that these questions are not going to be answered by
perturbation theory. To find the transition line  we return again to
MA and BDMC results.  In Fig. \ref{fig7}, we
begin with data similar to that displayed earlier for the SSH model,
i.e., plots of the variation of the quasiparticle behavior as a
function of $k$ and of $\lambda_{SSH}$, but now for a small but finite
$\lambda_H=0.25$. The results are shown here for the intermediate
adiabatic ratio $\Lambda_o=0.75$. We see that the curves are not
significantly changed from the pure SSH results by the addition of the
diagonal coupling. This is further confirmed by the data shown in
Fig. \ref{fig8}, for a larger diagonal coupling $\lambda_H=0.5$.

\begin{figure}[t]
\includegraphics[width=\columnwidth]{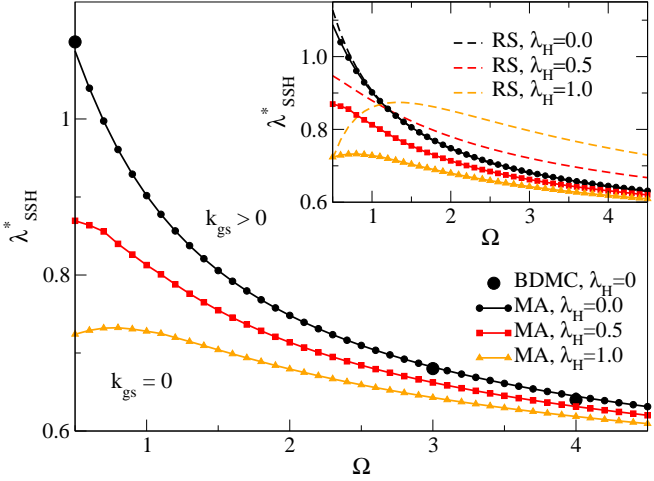}
\caption{\label{fig9} (color online) Transition coupling $\lambda^*_{SSH}$
  vs. $\Omega_o$. MA results are shown for $\lambda_H=0$ (circles),
  $\lambda_H=0.5$ (squares)
and $\lambda_H=1$ (triangles). Big circles show BDMC data for
$\lambda_H=0$. The inset compares these values with those predicted by
2$^{nd}$ order RS perturbation theory.
}
\end{figure}

Above we discussed two possible scenarios for the behavior of the
transition line as a function of the diagonal coupling. To see which
of these is actually enacted, we show in Fig. \ref{fig9}, the
transition value $\lambda^*_{SSH}$ as a function of $\lambda_H$. We
see that the critical line moves down as we increase $\lambda_H$, i.e.,
the diagonal coupling suppresses the bare nearest-neighbor coupling
more strongly than it does the higher nearest neighbor couplings.

Curiously, for larger $\lambda_H$ we see a peak in the critical line
followed by a decrease in $\lambda^*_{SSH}$ at small $\Omega_o$; this
is clear from the results for $\lambda_H = 1$, and the $\lambda_H=0.5$
case is also consistent with the possible existence of a peak at even
lower $\Omega_o$. This is reminiscent of the LPBED results found for
$\lambda_H=0$ at $\Lambda_o < 0.125$ (see Fig. 4 of
Ref. \onlinecite{PRL}). As before, we emphasize that our results are
not accurate in the limit of small $\Lambda_o$; however it is
interesting that they suggest that such a peak followed by a decrease
as $\Omega_o \rightarrow 0$ may actually be the typical behavior of
$\lambda^*_{SSH}$ in the adiabatic limit.

We also show, in the inset of Fig. \ref{fig9}, how the 2$^{nd}$-order
RS perturbation theory compares against the MA predictions. There is
no reason to expect perturbation theory to be accurate away from the
anti-adiabatic limit, and indeed that is the case for $\lambda_H\ne
0$; we believe that the excellent agreement for $\lambda_H=0$ is
accidental.

These results give only a partial characterization of the transition
behavior.  For a full characterization, we need a map of the {\it
  critical surface} $\mbox{\boldmath $\lambda$}^* (\Lambda_o)$, where
$\mbox{\boldmath $\lambda$}^* \equiv (\lambda_{SSH}^*, \lambda_H^*)$
is the locus of values of the 2-dimensional coupling $\mbox{\boldmath
  $\lambda$} = (\lambda_{SSH}, \lambda_H)$ upon which the effective
mass diverges.  Clearly a complete map of this critical surface is a
large undertaking, but just as one can produce a set of curves of
$\lambda^*_{SSH} (\Lambda_o)$ for different fixed values of
$\lambda_H$, one can also do the converse, i.e., produce a set of
curves of $\lambda^*_{H} (\Lambda_o)$ for different fixed values of
$\lambda_{SSH}$.

\begin{figure}[t]
\includegraphics[width=\columnwidth]{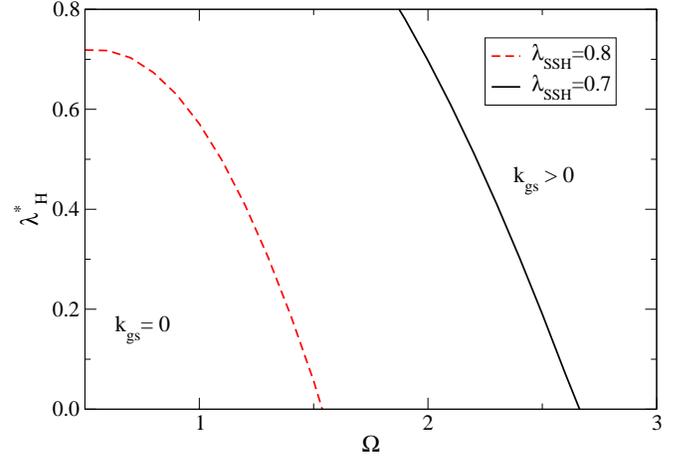}
\caption{\label{fig10} (color online) Transition coupling $\lambda^*_{H}$
  vs. $\Omega_o$ for the H/SSH model. MA results are shown for
  $\lambda_{SSH}=0.7$ (full
  line) and  $\lambda_{SSH}=0.8$ (dashed  line).}
\end{figure}

This is what Fig. \ref{fig10} does, showing two such curves for
$\lambda_{SSH}=0.7$ and $0.8$. Since the minimum value of
$\lambda^*_{SSH} (\Lambda_o)$ is $0.5$, as we reduce
$\lambda_{SSH}$ the region
occupied by the polaron dispersion with a $k_{gs}=0$ minimum should
grow, and when $\lambda_{SSH} \leq 0.5$ it should fill the
entire parameter space. Fig. \ref{fig10} confirms this expectation,
and we see that here a solution for a fixed $\lambda_{SSH}$ exists
only if $\Lambda_o$ is sufficiently small. This is unlike
Fig. \ref{fig9}, where a solution exists for a given $\lambda_H$ at
any $\Lambda_o$ (excluding, possibly, the strongly adiabatic region).

For $\lambda^*_H=0$ and $\lambda_{SSH}=0.8$ the
transition is at $\Omega_o \approx  1.53$, in agreement with
Fig. \ref{fig9}. As $\lambda^*_H$ is increased, the transition line
moves towards lower $\Omega_o$, enclosing the region where the
ground-state has the minimum at $k_{gs}=0$. As $\lambda_{SSH}$
decreases, this region indeed grows and should asymptotically fill the
entire phase-space.

This completes our detailed discussion of numerical and perturbative
results for the dual-coupling H/SSH model.

\section{An 'Effective Band' paradigm for polarons}

It is convenient to summarize all of these results in a new effective
theory for the polaron in its lowest band. We will see that this
picture is actually rather simple in the regime where $\Lambda_o =
\Omega_o/4t_o > 1$, i.e., where the phonons do not
overlap with the bare
band.

\subsection{Key features of the Dual Coupling Model}

Let us first consider the main features of our dual coupling model
results. We single out here the following as being most revealing:

\vspace{2mm}

(i) {\it Ground-state momentum}: In a purely diagonal coupling model,
nothing happens to the ground-state momentum - we always have $k_{gs}
= 0$. In the non-diagonal SSH model, $k_{gs}$ departs rapidly from
zero when we exceed the critical coupling $\lambda^*_{SSH}$ (itself a
function of $\Lambda_o$). In the dual coupling model, a non-zero
$k_{gs}$ is found everywhere above the critical surface
$\mbox{\boldmath $\lambda$}^* (\Lambda_o$).

\vspace{1mm}

(ii) {\it Ground-state effective mass}: In a purely diagonal model
like Holstein, the effective mass at the bottom of the lowest band
varies smoothly for any coupling strength - there is no 'band
collapse', only a rapidly decreasing bandwidth and rapidly increasing
effective mass, for large $\lambda_H$.  In the purely non-diagonal SSH
model, however, the mass does {\it not} become very large for large
coupling $\lambda_{SSH}$. Instead it continues to be of roughly the same
magnitude as the bare mass, {\it except} near the critical line
$\lambda^*_{SSH}(\Lambda_o)$, where it increases sharply, diverging to
infinity right on this line. For the dual coupling model, viewed as a
combined function of the dual coupling $\mbox{\boldmath
  $\lambda$}(\Lambda_o)$, we see these two behaviors combined - the
effective mass increases as before with $\lambda_{H}$, and for a given
fixed value of $\lambda_H$, it varies only slowly with changing
$\lambda_{SSH}$, except near the critical surface $\mbox{\boldmath
  $\lambda$}^* (\Lambda_o)$ where $m^*(k_{gs})$ diverges.

We emphasize yet again that no phase transition is involved here - all
that is happening is a change in the band shape, so that at a critical
value of the coupling, the band mass at {\it one specific value of
  momentum} diverges.

\vspace{1mm}

(iii) {\it Critical surface}: The position of
the critical surface depends only weakly on the adiabaticity parameter
$\Lambda_o$ for sections with constant $\lambda_H$, although varying
$\lambda_{SSH}$ can change things rapidly, as shown in
Fig. \ref{fig10}. Further work needs to be done to clarify this in the
adiabatic limit. We see that it is useful to think of the physics of
the dual-coupled model in terms of this critical surface. However,
this leaves unanswered the more fundamental question of what controls
the surface.

\vspace{1mm}

(iv) {\it Shape of Lowest Band}: This question is answered if
  we look at the bandshape. For diagonal coupling only, the shape of
  the lowest polaron band will not be strongly affected by the
  coupling to the phonons if $\Lambda_o = \Omega_o/4t_o >
  1$ (in one dimension). The band
  will narrow as we increase the coupling, and the $\cos k$ form will
  be weakly distorted by higher-order phonon terms, but $m^*(k=0)$
  will simply decrease smoothly as we increase the effective coupling,
  and the bottom of the band will remain at $k=0$. 

When we have non-diagonal couplings, large longer-range hoppings are
generated, which radically alter the band shape. This happens even
when $\Lambda_o \gg 1$ (cf. Fig. \ref{fig4}). Moreover, the overall
bandwidth does not become arbitrarily small with increasing effective coupling, like for diagonal coupling, but remains considerable. 

In the dual coupling model, we see that these two mechanisms work
relatively independently: the diagonal couplings cause a rapid but
smooth contraction of the bandwidth, with little change in the
bandshape, whereas the non-diagonal couplings alter the
bandshape and prevent the bandwidth from narrowing excessively.

\subsection{The effective band Hamiltonian}

Perhaps the most important insight to come from the studies recounted
here is that, provided the phonon excitations do not overlap the lowest
band (i.e., assuming $\Lambda_o > 1$), we can understand all of the
numerical results for the lowest polaron band in terms of a polaron
band energy $E_k$ characterized by a set of effective hopping
parameters $t_n (\mbox{\boldmath $\lambda$}, \Lambda_o)$ (compare
discussion in section IV.B, and Eq. (\ref{t-n})).

\subsubsection{Quasiparticle picture}

This suggests the following more general {\it ansatz} for the polaron
problem with both diagonal and non-diagonal couplings, when $\Lambda_o
> 1$. Suppose we start from a dual-coupling model having the general
form in Eq. (\ref{H-pol}), in which the phonons are assumed to be
optical in nature, i.e., with a gap energy $\Omega_o$. However, we
make no special assumptions about the form of $V(k,q)$, so that now we
have a starting Hamiltonian
\begin{eqnarray}
\hat{\mathcal{H}} &=& -\sum_{k} \epsilon_k ~ c_k^\dagger c_k + \sum_q
\Omega_o b_q^\dagger b_q \nonumber \\ & + & \frac{1}{\sqrt{N}}
\sum_{k,q} V(k,q) \; c_{k-q}^\dagger c_k \big( b_{q}^\dagger +b_{-q}
\big),
 \label{H-pol2}
\end{eqnarray}
where $\epsilon_k$ is the bare band dispersion (before coupling to
phonons).

Then we argue that for analysis of the low-energy polaron physics, we
can replace the Hamiltonian ${\cal H}$ in (\ref{H-pol2}) by an
effective Hamiltonian ${\cal H}_{\rm eff}$ having the general form
\begin{eqnarray}
\hat{\mathcal{H}}_{\rm eff} &=&\sum_{k}\tilde{\varepsilon}_k
\tilde{C}_k^\dagger \tilde{C}_k + \sum_q \tilde{\Omega}_o
\tilde{B}_q^\dagger \tilde{B}_q \nonumber \\ & + & \frac{1}{\sqrt{N}}
\sum_{k,q} \tilde{W}(k,q) \; \tilde{C}_{k-q}^\dagger \tilde{C}_k \big(
\tilde{B}_{q}^\dagger + \tilde{B}_{-q} \big),
 \label{H-pol3}
\end{eqnarray}
where 'tildes' denote renormalized quantities. As we saw previously,
if we start with a simple nearest-neighbor hopping in one dimension
(so $\epsilon_k = -2t_o \cos k$), the renormalized band energy takes
the form
\begin{equation}
\tilde{\varepsilon}_k^{(1d)} \;=\; -2\sum_n \tilde{t}_n
(\mbox{\boldmath $\lambda$}, \Lambda_o) \cos kn
 \label{effB}
\end{equation}
in which $\tilde{t}_1$ is the renormalized value of $t_o$, and the
$\tilde{t}_n$ are the set of new multi-site hopping parameters which
have been created by the non-diagonal coupling to phonons; the
operators $\tilde{C}_k$ create these new effective band
'quasiparticles'. The other terms in Eq. (\ref{H-pol3}) describe a set
of 'continuum' excitations, created by the bosonic operators
$\tilde{B}_q$, with renormalized energy $\tilde{\Omega}_o$, and a
renormalized 'hybridization' coupling $\tilde{W}(k,q)$ between the
band quasiparticles and the continuum excitations. There are new
energy scales in this effective Hamiltonian - a renormalized
interaction $\tilde{W}(k,q)$ between the effective band, with
renormalized bandwidth $\tilde{D} \sim \mathcal{O}(\tilde{t}_1)$, renormalized
gap $\tilde{\Omega}_o$ to continuum excitations, and a renormalized
adiabaticity parameter $\tilde{\Lambda}_o =
\tilde{\Omega}_o/\tilde{D}$.

The key advantage of this quasiparticle picture is that except in the
{\it effective} anti-adiabatic regime, where the effective
adiabaticity parameter $\tilde{\Lambda}_o < 1$, the renormalized
coupling $\tilde{W}(k,q)$ plays little role - it is considerably
smaller than the original $V(k,q)$, most of whose effects have already
been absorbed into the renormalized band. What this means is that
unless $\tilde{\Lambda}_o \sim \mathcal{O}(1)$ or less, the effective band
energy $\tilde{\varepsilon}_k$ will differ very little from the exact
$E_k$ that one finds in either an exact numerical calculation, or in
experiments. Thus we will find that
\begin{equation}
\tilde{t}_n \sim t_n \;\;\;\;\;\;\;\;(\tilde{\Lambda}_o > 1)
 \label{t-n2}
\end{equation}
where the coefficients $\{ t_n \}$ are extracted from fitting to the
exact band dispersion $E_k$ (or numerical approximations to it). Only
when, as $\tilde{\Lambda}_o$ is decreased, the continuum states
approach the top of the effective band, will there be significant
differences between $\tilde{\varepsilon}_k$ and $E_k$, because of
level repulsion between the effective band and the continuum
excitations. The quasiparticle picture discussed here assumes that
$\tilde{\Lambda_o} > 1$ is always satisfied.

We make no attempt here to determine the parameters in this effective
Hamiltonian as functions of the underlying parameters in Eq.
(\ref{H-pol}). The philosophy we adopt is similar to that
involved in,  eg., the Landau Fermi liquid theory: the parameters in
$\hat{\mathcal{H}}_{\rm eff}$ are ultimately to be determined by
experiment on specific system(s), where possible; one predicts the
results for different experiments in terms of the $\tilde{t}_n$, and
uses the results to determine the $\tilde{t}_n$.

Thus the form given above for $\hat{\mathcal{H}}_{\rm eff}$
constitutes a hypothesis for the results of both experiments and
non-perturbative numerical work, and is thus in principle testable.
The key result in the present work, as we saw in section IV.B, is that
all of the features found here in the dual coupling model can be
understood in terms of the behavior of the $\{ t_n \}$ defined in
(\ref{t-n}); the following aspects are particularly important:

\vspace{1mm}

(a) If we only have diagonal couplings in the model, then only
$\tilde{t}_1$ exists in the sum of Eq. (\ref{effB}) and we have a very
simple band polaron, whose bandwidth and effective mass vary
continuously with coupling strength. The actual dispersion $E_k$ will
hardly differ from the quasiparticle form $\tilde{\varepsilon}_k =
-2\tilde{t}_1 \cos k$ (in one dimension) when $\tilde{\Lambda}_o$ is
large; only when the continuum gap energy $\tilde{\Omega}_o\sim
\tilde{t}_1$ will we see a distortion of the simple cosine form for
the band dispersion. To see this, consider Fig. \ref{fig1} (c), where
$\lambda_H = 1$ in the Holstein model; the distortion of $E_k$ away
from a simple cosine form, caused by the continuum states, is quite
small even when $\Lambda_o$ is as small as $0.5$. This is because here
$\tilde{t}_1< 0.15t_o$, so that the
renormalized adiabaticity parameter $\tilde{\Lambda}_o \sim 1$; the
continuum is well  above the lower quasiparticle
band. For large diagonal coupling, the $\cos k$ form for the
quasiparticle band will be accurate except for very small $\Omega_o$.

\vspace{1mm}

(b) Introducing a non-diagonal coupling generates the $\tilde{t}_n$
with $n=2,3,...$; with increased coupling strength, they become
steadily more important.  If their signs are negative, a
  transition to a new polaron with finite $k_{gs}$ is eventually
  achieved, once $|\tilde{t}_2|$ outweighs $\tilde{t}_1$; as
$|\tilde{t}_2|/\tilde{t}_1$ increases further, $k_{gs} \rightarrow
\pi/2$. A key result comes out of the H/SSH model, where we saw that
increasing the diagonal coupling $\lambda_H$, while keeping the
non-diagonal coupling $\lambda_{SSH}$ constant, suppresses the
lowest-order hopping term $t_1$ faster than it does the higher terms
$t_2, t_3$, etc. This then makes it easier for the system to make the
transition (i.e., it then happens for a smaller value of
$\lambda_{SSH}$). The renormalized band, now parametrized in terms of
the $\{ \tilde{t}_n \}$ simply evolves in shape as $\lambda_{SSH}$
increases. For any finite diagonal $\lambda_H$, the bandwidth remains
finite, no matter what is the non-diagonal $\lambda_{SSH}$.

\vspace{1mm}

(c) The transition line in no way marks any kind of quantum critical
point, or collapse of the bandwidth. It simply marks the locus of
points where the ground state momentum starts to shift away from zero.
The renormalized bandwidth is never zero (see the plots of $E_k$ in
Figs. \ref{fig2}-\ref{fig4}, \ref{fig7}, and \ref{fig8}); although the
$k=0$ effective mass diverges along the transition line, this is
simply because $k=0$ then marks a point of inflection of the
dispersion relation $E_k$, at which $\partial^2 E_k/\partial k^2 = 0$. The
renormalized adiabaticity ratio $\tilde{\Lambda}_o$ also is
finite at the transition point.

\subsubsection{Transitions in the band structure}

We have already remarked above that the transition found here is
completely unconnected with the rapid crossover found in the polaronic
bandwidth as $\lambda_H$ is increased - indeed there is never a
transition as a function of $\lambda_H$. It also seems unlikely that
anything in particular should happen to the electron-phonon
entanglement at the transition we have found, since the effective band
structure is varying continuously, and certainly for $\Lambda_o \gg
1$, nothing special happens to the mixing between the polaronic band
and the higher energy continuum. In this connection we note that Ref.
\onlinecite{stojan08} finds some sort of transition in the
entanglement between the polaron and the phonons, however we believe
that what they have found is unconnected to the transition discussed
here. This is because unlike what we find, their transition only
occurs for sufficiently small $\Lambda_o$, and even when present it is
at different values of the coupling.

We emphasize that one expects to see rather complicated effects when
the lowest polaron quasiparticle band begins to overlap with the
optical phonon - at this point level-crossing and real transitions
become possible, with a large scale reorganization of states at the
top of the polaron band. This is not a quantum phase transition, at
least in the conventional sense of the term (indeed, if one applied
the term to describe this overlap of states for a single
quasiparticle, it could also be applied to describe a wide variety of
chemical and nuclear reactions, which would completely divest it of
its original significance).

\section{Conclusions: More General Models, and Experiments}

All the concrete results found in sections IV and V were for the dual
H/SSH model. However this model is still fairly specialized; it
involves only couplings linear in the phonon variables, and couples
electrons to gapped optical phonons only. How general are such
results, and what might they have to do with experiments on real
systems? Let us look at these questions in turn.

\subsection{More general couplings}

We can obviously ask what happens if (i) we go to
more general models with combined diagonal and non-diagonal couplings,
and (ii) if we also include acoustic phonons, and even perhaps
higher-order phonon terms, such as the quadratic couplings
in Eq. (\ref{diag}).

Once we look at the system in terms of the effective band Hamiltonian,
it becomes clear that there is almost certainly a lot more interesting
physics to be revealed in dual coupling models of this kind. This is
because there is no reason in principle why one cannot imagine a
destabilization of the $k_{gs} = 0$ groundstate to other ground states
in which, for example, the $t_3$ hopping term dominates (which could,
as $t_3$ began to dominate over $t_1$, have a $k_{gs} \rightarrow
\pi/3$ or $2\pi/3$). In this way one can envisage a hierarchy of
states, each having a different groundstate momentum, with critical
lines/surfaces existing between them. Clearly, to see this full
complexity, one would have to begin by allowing modifications of the
form of the diagonal and non-diagonal couplings $g_1({ q}), g_2({ k},{
  q})$, thereby exploiting the full freedom allowed by the momentum
dependence of each - the Holstein and SSH coupling forms are merely
the simplest that one can imagine. One way to get more interesting
physics of this kind would be looking for non-diagonal couplings
$g_2({ k}, { q})$ with longer-range in real space, i.e., with a more
sharply-peaked behavior in momentum space. Thus a fairly clear
prediction arising from our effective Hamiltonian is that of possible
multiple "effective band" transitions as one varies the non-diagonal
couplings.

Another possibility, recently demonstrated in a model combining SSH
and BM coupling,\cite{MAca} appears when the diagonal coupling is also
longer-range. In such cases, interference between the two couplings
can lead to additional renormalization of $t_o$ and even change its
sign. For example, in the limit $\Lambda_o \gg 1$ of Ref.
\onlinecite{MAca}, this occurs through processes where SSH coupling
moves the electron to a neighboring site while emitting a phonon at the
original site, and then BM coupling absorbs that phonon without
changing the electron location (the order of the two processes can be
reversed and the contributions add up). This generates an additional
contribution to $t_1$ whose sign is controlled by the product of the
two couplings. If it is negative and if its magnitude is large
enough, one expects to see a second transition to a $k_{gs} = \pi$. Indeed,
this was observed in Ref. \onlinecite{MAca}.

The way in which any coupling to acoustic phonons and phonon pair
excitations might affect these results is an interesting question. We
tend to believe that they will only further renormalize the effective
band energy $\tilde{\varepsilon}_k$, as well as introduce dissipative
processes for arbitrarily low energy, but that they will not alter the
basic result that non-diagonal coupling to optical phonons can
reorganize the band and change the ground-state momentum. However, this
question merits further study.

Finally, it is now known that even small non-linear coupling to
phonons, of the Holstein type (i.e. local in real space) has
significant effects both on the physics of the single
polaron,\cite{clemens} and at finite carrier
concentrations.\cite{steve} It should be clear from the above
discussion that such non-linear terms in either kind of coupling may open
additional channels to generate longer-range hoppings. Whether
this can lead to qualitatively different behavior from that discussed
above for linear couplings is another question that merits further
study.

\subsection{Experiments and Conclusions}

So far, we have not discussed the experimental relevance
  of our work. The main reason is that, apart from a few notable
  exceptions,\cite{troisi07} both computational efforts to calculate
  these couplings and the models used to extract such couplings from
fits of  experimental measurements tend to focus on a pure model, {\em i.e.}
  either on only diagonal or only non-diagonal coupling.  What would
be far preferable would be attempts to compare theory and experiment
directly, using the kind of computations discussed here - experiments would
allow one to extract the renormalized parameters rather than the bare
ones, and they will be quite different if the couplings are strong. It
almost goes without saying, in view of the results found here, that
any attempt to compare experiments with theories which do not include
the effect of Peierls couplings is likely to give very misleading
results.

 A different avenue is opened by the use of cold atoms
  and/or polar molecules trapped in optical lattices as simulators of
  such mixed Hamiltonians.\cite{Herrera,MAca,RK1,RK2} As typical in such
  systems, they would permit the tuning of individual coupling
  strengths within wide ranges, such that properties like those
  discussed here could be investigated in a significant area of the
  parameter space.

To conclude, we have given a fairly exhaustive characterization of the
behavior of polarons in the dual coupling H/SSH model, in terms of an
'effective band' paradigm, in which diagonal interactions cause a
continuous narrowing of a low-energy quasiparticle band, and
non-diagonal interactions add non-local hopping terms to this band,
which gradually change its shape, and eventually force the ground
state to move, at a definite transition point, to non-zero momentum.
Indeed, we see no evidence for, nor any theoretical reason
to expect, any sudden changes to the polaronic band, for any value of
either diagonal or non-diagonal coupling to phonons. The effective
band quasiparticle picture works if the phonon energy is larger than
the renormalized bandwidth - otherwise level mixing between the
phonons and the polaronic states gives a more complicated picture. The
behavior of physical quantities in an H/SSH model will typically be
rather different from what one expects from a simple Holstein model,
and this needs to be taken into account in comparison between theory
and experiment.

\acknowledgements
We would like to thank Prof N.V. Prokof'ev for discussions of this
work. DJJM was supported by NSERC, PCES and MB by NSERC and CIFAR, and
PCES also by PITP.

\appendix

\section{Bold Diagrammatic Monte Carlo technique}

We herein distinguish between methods
sampling the Green's function ($G$-DMC), and the self-energy
($\Sigma$-DMC). The Bold Diagrammatic Monte Carlo (BDMC) algorithm is
essentially a $\Sigma$-DMC method, but relies on a faster sampling
based on a self-consistent procedure relying on Dyson's identity \cite{PhysRevLett.99.250201,PhysRevB.77.020408,PhysRevB.77.125101}.
In this appendix, we provide a more detailed summary of our own implementation to polaron problems.

\subsection{Green function, self-energy and diagrammatics}

The main quantity of interest here is the $T=0$
retarded polaron Green function. We write this as
\begin{equation}
 \label{eq_g_real_time}
G(t,\, k,\, \mu)=-i\Theta(t) \sum_n e^{-it\big(E_n (k)-\mu\big)}
|\langle n,\, k | k\rangle|^2,
\end{equation}
in terms of eigenstates $|n,\,k\rangle$ and $E_n(k)$ of the full Hamiltonian; here $\mu$ is the
chemical potential, which for this one-electron system is just an
overall shift in energy. In the absence of interactions this reduces to
$G_0(t,\,k,\,\mu)=-i\Theta(t) e^{-it\big(\epsilon (k)-\mu\big)}$. In the frequency domain we write
\begin{equation}
\label{eq_g_real_frequency}
G(\omega,\, k,\, \mu) = \sum_n \frac{Z_n(k)}{\omega - E_n (k) +\mu
+i\eta},
\end{equation}
with infinitesimal $\eta > 0$.

\begin{figure}[bt]
\begin{center}
\includegraphics[width=1\columnwidth]{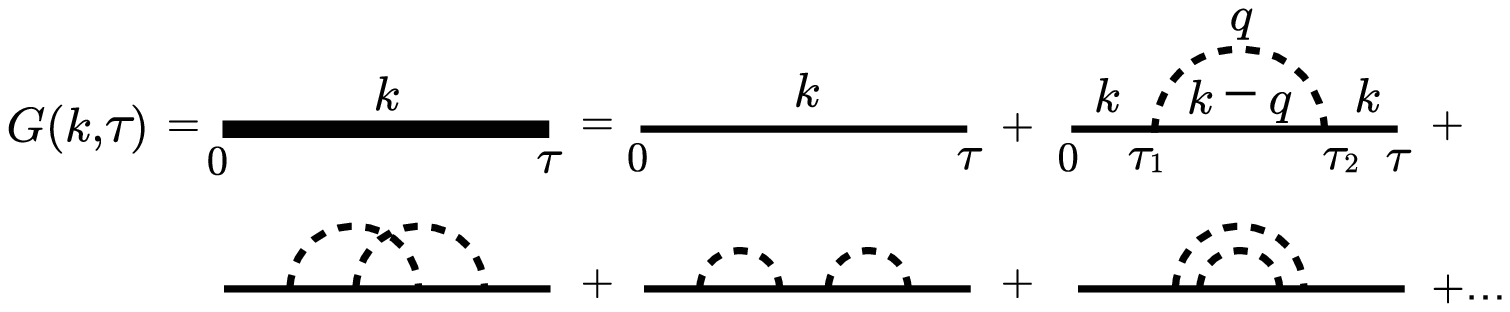}
\caption{\label{fig11}Diagrams contributing to
$G^{i,\,1^\textrm{\tiny st}}(\tau,\,k)$. Relevant indices are only shown for the the
first few diagrams. Each phonon propagator $i$ has a momentum $q_i$,
and each electron propagator $j$ has a momentum
$k_j=k-\sum_{i'\in\{j\}_{i'}} q_{i'}$ where $\{j\}_{i'}$ is the set of phonons
propagators covering the electron propagator $i'$}
\end{center}
\end{figure}

As noted by Prokof'ev et al., the sign problem can be avoided by
analytic continuation of the Green's function to imaginary time $\tau=it$, and corresponding imaginary frequency $\xi=-i\omega$, so that the exponential in (\ref{eq_g_real_time}) is
real and positive definite. However,
the analytic continuation and the Fourier transformation do not
commute. Writing  \cite{PhysRevB.77.020408}
$G^i ( \xi,\,k,\, \mu)=G (\omega=i\xi,\, k,\, \mu)$,
we define the imaginary time Green's function to be
\begin{equation}
G^i (\tau,\, k,\, \mu) = -i G(t=-i\tau,\, k,\, \mu),
\end{equation}
where the superscript $i$ denotes imaginary quantities. For convenience
we actually calculate $\tilde{G}^i(\tau,\,k,\,\mu)=-G^i(\tau,\,k,\,\mu)$. Fig. \ref{fig11} shows $G^i(k,\,\tau)$ up to second order diagrams. The relevant
indices which need to be sampled are shown only for the first order diagram.

For $\Sigma$-DMC and BDMC we need to use the proper
self-energy, defined as usual by organizing $G^i$ diagrams as blocks
of inseparable phonon tangles linked by bare electron propagators, so that the self-energy is the sum of all possible diagrams that cannot be separated into a number of pieces
by cutting only an electron line. Then Dyson's equation
in imaginary frequency reads
\begin{equation}\label{eqn_g_imaginary_frequency_dyson}
G^i(\xi,\,k,\,\mu) = \frac{1}{G^{i\:\:-1}_0(\xi,\,k,\,\mu) -
\Sigma^i(\xi,\,k,\,\mu)},
\end{equation}
where $\Sigma^i(\xi,\,k,\,\mu)$ is the Fourier transform of
$\Sigma^i(\tau,\,k,\,\mu)$. Prokof'ev and Svistunov showed \cite{PhysRevB.77.125101} that most
of the relevant observables can be extracted directly from the
imaginary time self-energy without using Dyson's equation
first to obtain the Green's function; $\Sigma^i(\xi,\,k,\,\mu)$ is calculated with the DMC
algorithm using similar diagrammatic rules to those of $G$-DMC for propagators and
the interaction vertices, but now using the topology of the $\Sigma$ diagrams. Each diagram included in $\Sigma$ corresponds to an
infinite number of diagrams in $G$, making  $\Sigma$-DMC much more efficient.

\subsection{MC Sampling, normalization and orthonormal functions}

In both $G$-DMC and $\Sigma$-DMC. it is useful to introduce a
constant shift in energy $\mu$ to allow for a good sampling of diagrams with larger imaginary time. 
In both $G$-DMC and $\Sigma$-DMC. In $G$-DMC, normalization is most easily achieved using
$\tilde{G}^i(\tau=0,\, k)=1$. We similarly can normalize
$\tilde{\Sigma}^i(\tau,\,k,\,\mu)=-\Sigma^i(\tau,\,k,\,\mu)$ using
\begin{equation}\label{eqn_norm_se}
\tilde{\Sigma}^i (\tau=0,\,k)=\frac{1}{2\pi} \int_{0}^{2\pi}
\!\!\textrm{d}q\, g(k-q,\,q) g(k,\,-q),
\end{equation}
which differs from $\tilde{G}^i(\tau=0,\, k)$ due to the contribution of
two vertices. This is the approach used in this work. Alternative
normalization schemes consist in enlarging the configuration space with
unphysical diagrams that can be calculated analytically or selecting a
subset of the physical diagrams that can be calculated analytically. The
other diagrams are then normalized by calculating the ratio between this
normalization sector and the rest of the configuration space.

The accuracy of the normalization and of the data collected will
depend closely on the histogram spacing used. A notable improvement
over a simple histogram is to use a bin size increasing
with $\tau$, giving a better sampling at small times for an
accurate normalization, while accumulating enough points per
bin at large $\tau$ to average over statistical noise. When collecting
statistics with variable-size bins, we need to include a factor of
$1/\delta \tau$, with $\delta \tau$ the size of the bin.

A further improvement consists in expanding the function sampled over
the range of each bin with a set of orthonormal functions
$F_n(\tau-\tau_m)$ centred at the bin centre $\tau_m$. An arbitrary
function $f(\tau)$ can be written as
\begin{equation}\label{eqn_fn_orthonormal_reconstruction}
f(\tau) = \sum_a c_a F_a(\tau-\tau_m) + \textrm {correction},
\end{equation}
for $\tau \in [\tau_m - \delta\tau/2,\, \tau_m + \delta\tau/2]$, with coefficients
$c_a$. A finite set of
functions is sufficient if the bin size $\delta\tau$ is not too
large, and if the function is smooth on this range. The Gram-Schmidt
orthogonalization procedure can be used to normalize the chosen set.
The functions used herein are the Legendre polynomials
$P_n(\tau-\tau_m)$, but normalized such that
$\int_{-\delta\tau/2}^{\delta\tau/2}\textrm{d}\tau F_a^2(\tau) = 1$,
instead of the usual $P_n(1)=1$. Collecting statistics for a specific bin
is now done for each coefficient $c_a$. After an update, if the diagram
has a length that falls on the bin's range, each coefficient is updated with
\begin{equation}
c_a \to c_a + F_a(\tau-\tau_m).
\end{equation}

\subsection{Updates}

The minimal set of updates is similar to $G$-DMC. The set must 
include updates to insert and delete phonons,
and an update to change the total length of the self-energy diagram. We
must start with at least one phonon and we cannot allow this first
phonon to be removed; so an extra update to change the
momentum of this first phonon is needed. The absence of the
bare propagator before and after the proper self-energy part means that the
insertion of a phonon must now allow for a change of the diagram
length by considering also phonons inserted past the current length of
the diagram. Each type of update is assumed to be chosen with equal
probability $1/4$. We refer below to the current diagram with all its
specific values for time and momentum indices as $\nu$, and to the
suggested updated diagram as $\nu'$, while $W_\nu$ ($W_{\nu'}$) are the weights
of the current (updated) diagram.

Phonon insertion and deletion can be
balanced together. For insertion, we sample the
momentum uniformly over $[0,\,2\pi[$, then choose the time of the
first vertex $\tau_1$ uniformly on $]0,\,\tau]$ with $\tau$ the time
of the final vertex of the diagram $\nu$ . We
use the value of the phonon propagator $\propto e^{-\omega(q)(\tau_2
-\tau_1)}$ to sample the time of the second vertex between
$[\tau_1,\,\tau_\textrm{\tiny max}]$, where $\tau_\textrm{\tiny max}$
is some fixed cutoff. This keeps the probability distribution closer
to the ratio of weights, but ignores the more complicated values of
the electron propagators. The exponential distribution also prevents
too many update suggestions that would result in very long and costly
diagrams. The resulting probability distribution for inserting a
specific phonon is then
\begin{equation}
\mathcal{P}_\textrm{\small insert} =
\frac{1}{4}\cdot\frac{1}{2\pi}\cdot\frac{1}{\tau}\cdot\frac{\omega(q)
e^{-\omega(q)(\tau_2-\tau_1)}}{1-e^{-\omega(q)(\tau_\textrm{\tiny
max}-\tau_1)}}.
\end{equation}
The inverse update, deletion, simply selects one phonon out of the
$n_\textrm{\tiny ph}$ phonons not including the first one, with a
probability of
\begin{equation}
\mathcal{P}_\textrm{\small delete} =
\frac{1}{4}\cdot\frac{1}{n_\textrm{\tiny ph}-1}.
\end{equation}
The acceptance ratios for the updates are therefore
\begin{equation}
R_\textrm{\small insert} = \mathcal{F}(\nu')\cdot
\frac{W_{\nu'}}{W_{\nu}} \frac{2\pi \tau}{n_\textrm{\tiny
ph}}\cdot\frac{1-e^{-\omega(q)(\tau_\textrm{\tiny
max}-\tau_1)}}{\omega(q) e^{-\omega(q)(\tau_2-\tau_1)}},
\end{equation}
\begin{equation}
R_\textrm{\small delete} = \mathcal{F}(\nu')\cdot
\frac{W_{\nu'}}{W_{\nu}} \frac{(n_\textrm{\tiny
ph}-1)}{2\pi\tau}\cdot\frac{\omega(q)
e^{-\omega(q)(\tau_2-\tau_1)}}{1-e^{-\omega(q)(\tau_\textrm{\tiny
max}-\tau_1)}},
\end{equation}
where the extra factor
\begin{eqnarray}
\mathcal{F}(\nu') &= & \:1 \quad \textrm{if $\nu'$ is a proper
$\Sigma$ diagram}\nonumber\\ & & \:0 \quad \textrm{otherwise},
\end{eqnarray}
prevents improper diagrams from being generated.

In our own implementation we actually replace the update to change the
length of the diagram by a more general update that will shift any
vertex to a new time. We first select one of the vertices at random, except for
the first one which we will keep fixed at $\tau=0$, and then select a
new position for this vertex anywhere between the previous vertex and
the next one. The last vertex is a special case for which we will
select a new position between the previous one and a maximum
$\tau_\textrm{\tiny max}$. In its simplest version, the position is
uniformly distributed such that the acceptance ratio is simply the ratio of the weights.

Finally the update to change the momentum of a phonon simply selects
one of the phonons and changes the momentum from its previous value to
any other value with equal probability, again with an acceptance ratio equal to the
ratio of the weights.

Given that we never remove the first phonon, ergodicity for the special case of a
Hamiltonian with more than one branch will require an extra update to change
the phonon branch of a specific phonon.

\subsection{BDMC, bold line and double-counting}

\begin{figure}[bt]
\begin{center}
\includegraphics[width=1\columnwidth]{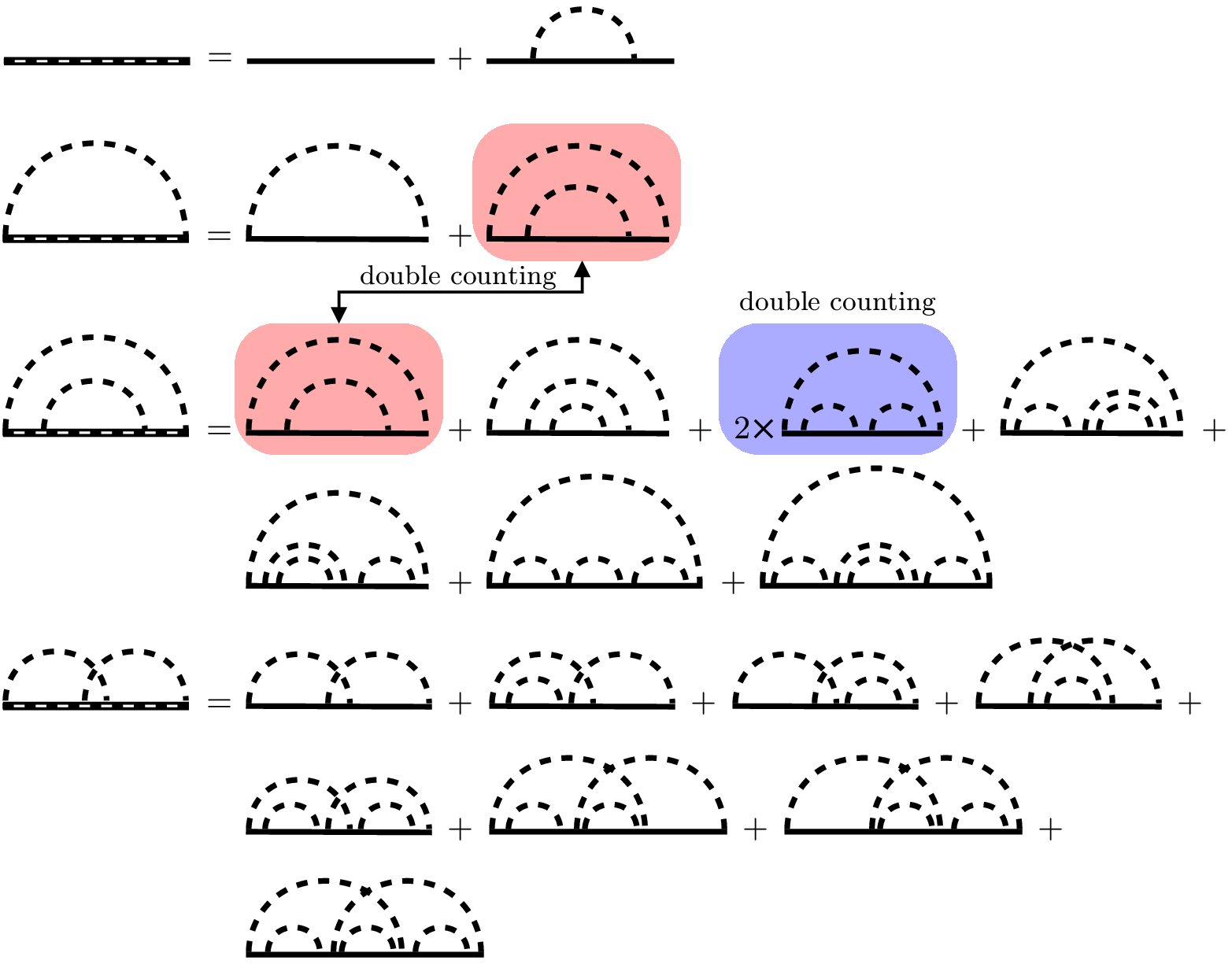}
\caption{\label{fig13}Drawing first and second
order self-energy diagrams with a more complicated (bold) propagator composed
of the bare propagator and first order Green's function diagrams. The bold propagator
is shown as a dashed thick line. The diagrams are assumed to be in frequency space
such that the length of a propagator does not matter.  }
\end{center}
\end{figure}

Fig. \ref{fig13} shows how we implement the BDMC algorithm. 
The first-order diagrams for the self-energy generated with the thick "bold"
line give two diagrams while the two second-order diagrams now account
for 15 types of diagrams. We have, however, double-counted 2 types of
diagrams up to this order, something that needs to be forbidden to obtain the
correct answer. To improve the bold propagator self-consistently the BDMC
algorithm goes as follows:

\begin{noindlist}
\item Initialize $\tilde{G}^i_\approx
(\tau,\,k,\,\mu)=\tilde{G}^i_0(\tau,\,k,\,\mu)$ for a set $\{k\}$.
\item \noindent Draw a first diagram for each $k$.
\item MC sampling of diagrams for $\tilde{\Sigma}^i(\tau,\,k,\,\mu)$ for
each $k$. Diagrams are drawn using $\tilde{G}^i_\approx$. Repeat $n$ times.
\item Fourier transform $\tilde{\Sigma}^i(\tau,\,k,\,\mu)$ to get
$\tilde{\Sigma}^i(\xi,\,k,\,\mu)$.
\item Use Dyson equation to get
$\tilde{G}^i_\approx (\xi,\,k,\,\mu)$.
\item Fourier transform back to get a new $\tilde{G}^i_\approx (\tau,\,k,\,\mu)$.
\item Go back to step 3 until convergence $\tilde{G}^i_\approx \approx \tilde{G}^i$
\end{noindlist}

We note the need to solve for all momenta at once because the momentum
of a specific electron propagator can have any value after a phonon is
created. The chemical potential should also be made momentum-dependent, to
ensure that for each value of momentum calculated, the method samples the large $\tau$
behavior accurately. When calculating the bold propagator of momentum
$k'$ inside a diagram of momentum $k$, we will need to match the
chemical potential by adjusting $\tilde{G}^i_\approx\big(\tau,\, k',\, \mu(k')\big)$ to $\tilde{G}^i_\approx\big(\tau,\, k',\, \mu(k')\big)\cdot
\exp\big([\mu(k)-\mu(k')]\tau\big)$ to calculate the value of the
diagram. The set of updates for $\Sigma$-DMC described above only requires
one modification to $\mathcal{F}(\nu')$ which should now return 1
only if the diagram is not going to cause any double counting, and $0$
otherwise.

\begin{figure}[hbt]
\begin{center}
\includegraphics[width=1\columnwidth]{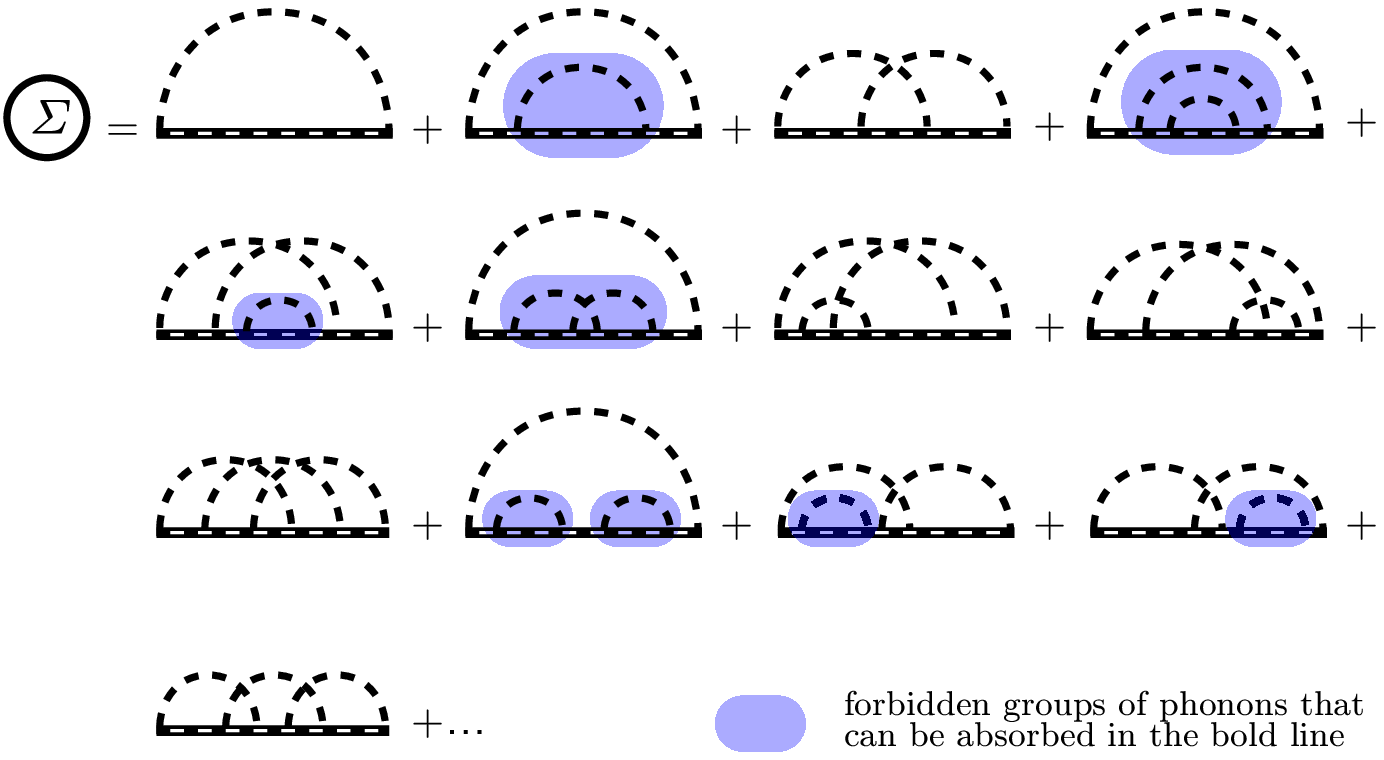}
\caption[Forbidden and allowed self-energy bold diagrams up to third
order]{\label{fig14}Forbidden and
allowed self-energy bold diagrams up to third order}
\end{center}
\end{figure}

\begin{figure}[hbt]
\begin{center}
\includegraphics[width=1\columnwidth]{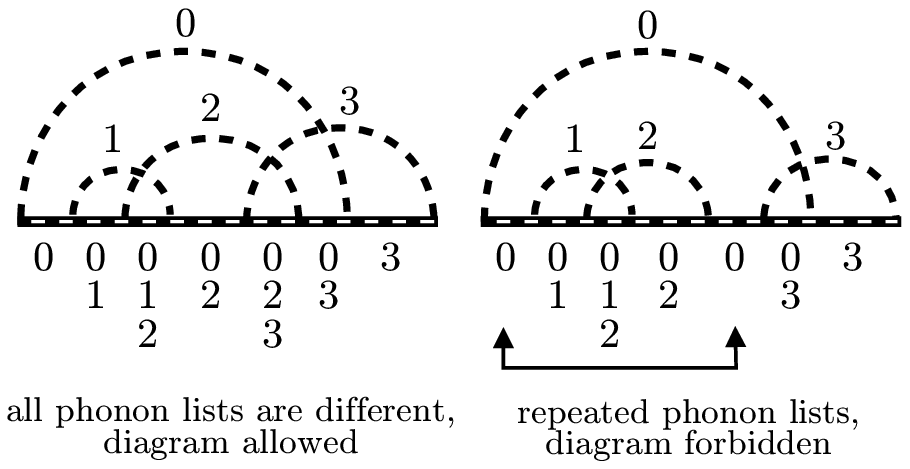}
\caption[Algorithm to check if a self-energy diagram is allowed in
BDMC]{\label{fig15}Algorithm to check
if a self-energy diagram is allowed in BDMC. Each phonon is assigned a
unique number and each propagator has a list of phonons covering
it. Diagrams with each list being unique are allowed, and diagrams
with repeated lists are forbidden.}
\end{center}
\end{figure}

Since we are now using a bold line, we need to avoid drawing a diagram
which could be obtained by expanding the bold line. In other words,
any phonon or group of self-contained phonons lines that can be
absorbed in the bold line without changing the topology and the
momentum of the rest of the diagram should not be allowed. By a
self-contained group of phonons, we mean that the group only has one
incoming and one outgoing electron line of the same momentum, and no
other phonon line. Fig. \ref{fig14}
gives a few examples of forbidden and allowed diagrams. Fig.
\ref{fig15} shows a simple algorithm to
check if a diagram is forbidden or allowed by assigning each phonon a
unique number and creating lists of phonons covering each electron
propagator. Each propagator needs to have a distinct phonon list to be
allowed. Allowed diagrams are referred to as \emph{fully crossed
diagrams}.

\begin{thebibliography}{99}
\bibitem{divergentS}  It is important that in the DQMC and BDMC methods, calculations
                      are done over a finite time interval, which is roughly equivalent to introducing an IR cut-off. If this were not done, then high-order terms in the perturbation series would eventually dominate - the series is of course divergent - and the series would 'blow up' at high orders. At finite time $t$, a factor $t^{2n}/(2n)!$ more than compensates for the number of diagrams $\sim 2^n n^{3/2} n!$ containing $n$ phonon lines.
\end{thebibliography}
\end{document}